\documentstyle[12pt,equations,epsf]{article}
\newcommand{\be}{\begin{equation}}
\newcommand{\ee}{\end{equation}}
\newcommand{\bea}{\begin{eqnarray}}
\newcommand{\eea}{\end{eqnarray}}
\newcommand{\nn}{\nonumber}

\begin{document}

\def\gamh{\Gamma_H}
\def\eb{E_{\rm beam}}
\def\deb{\Delta E_{\rm beam}}
\def\sigm{\sigma_M}
\def\sigmmax{\sigma_M^{\rm max}}
\def\sigmmin{\sigma_M^{\rm min}}
\def\sige{\sigma_E}
\def\dsigm{\Delta\sigma_M}
\def\mh{M_H}
\def\lyear{L_{\rm year}}

\def\wstar{W^\star}
\def\zstar{Z^\star}
\def\ie{{\it i.e.}}
\def\etal{{\it et al.}}
\def\eg{{\it e.g.}}
\def\pzero{P^0}
\def\mt{m_t}
\def\mpzero{M_{\pzero}}
\def\mev{~{\rm MeV}}
\def\gev{~{\rm GeV}}
\def\gam{\gamma}
\def\lsim{\mathrel{\raise.3ex\hbox{$<$\kern-.75em\lower1ex\hbox{$\sim$}}}}
\def\gsim{\mathrel{\raise.3ex\hbox{$>$\kern-.75em\lower1ex\hbox{$\sim$}}}}
\def\ntc{N_{TC}}
\def\epem{e^+e^-}
\def\tauptaum{\tau^+\tau^-}
\def\lplm{\ell^+\ell^-}
\def\anti{\overline}
\def\mw{M_W}
\def\mz{M_Z}
\def\fbi{~{\rm fb}^{-1}}
\def\mupmum{\mu^+\mu^-}
\def\rts{\sqrt s}
\def\sigrts{\sigma_{\tiny\rts}^{}}
\def\sigrtssq{\sigma_{\tiny\rts}^2}
\def\sigrtsprime{\sigma_{E}}
\def\nsigrts{n_{\sigrts}}
\def\gampzero{\Gamma_{\pzero}}
\def\pzerop{P^{0\,\prime}}
\def\mpzerop{M_{\pzerop}}

\font\fortssbx=cmssbx10 scaled \magstep2
\hbox to \hsize{
%
%
$\vcenter{
\hbox{\fortssbx University of California - Davis}
\hbox{\fortssbx University of Florence}
\hbox{\fortssbx University of Geneva}
}$
\hfill
$\vcenter{
\hbox{\bf UCD-99-5}
\hbox{\bf DFF-333/2/99}
\hbox{\bf UGVA-DPT-1999 03-1029}
\hbox{April, 1999}
}$
}

%
\medskip
\begin{center}

{\Large\bf\boldmath Analysis of Narrow $s$-channel Resonances
at Lepton Colliders\\}
\rm
\vskip1pc
{\Large R. Casalbuoni$^{a,b}$, A. Deandrea$^c$, S. De Curtis$^b$,
\\ D. Dominici$^{a,b}$, R. Gatto$^d$ and J. F. Gunion$^e$\\}
\vspace{5mm}
{\it{$^a$Dipartimento di Fisica, Universit\`a di Firenze, I-50125
Firenze, Italia
\\
$^b$I.N.F.N., Sezione di Firenze, I-50125 Firenze, Italia\\
$^c$Institut Theor. Physik,  Heidelberg University, D-69120
             Heidelberg, Germany
\\ $^d$D\'epart. de Physique Th\'eorique, Universit\'e de
Gen\`eve, CH-1211 Gen\`eve 4, Suisse
\\
$^e$Department of Physics, University of California,
Davis, CA 95616, USA}}
\end{center}
\bigskip
\begin{abstract}

\noindent
The procedures for studying a single narrow $s$-channel
resonance or nearly degenerate resonances at a lepton collider, especially
a muon collider, are discussed. In particular, we examine four methods for
determining the parameters of a narrow $s$-channel resonance: scanning
the resonance, measuring  the convoluted cross section, measuring
the Breit-Wigner area, and sitting on the resonance
while varying the beam energy resolution. This latter
procedure is new and appears to be potentially very powerful.
Our focus is on computing the errors in resonance parameters
resulting from uncertainty in the beam energy spread. Means
for minimizing these errors are discussed.
The discussion is applied to the examples of a
light SM-Higgs, of the lightest pseudogoldstone boson
of strong electroweak breaking,
and of the two spin-1 resonances of the Degenerate BESS model (assuming
that the beam energy spread is less
than their mass splitting). We also examine the most effective procedures
for nearly degenerate resonances, and apply these to the case of
Degenerate BESS resonances with mass splitting of order the beam energy spread.
\end{abstract}

\section{Introduction}
\bigskip
The possibility of analyzing narrow $s$-channel
resonances is considered to be one of the most important strengths
of muon colliders,  at present under consideration by different laboratories
\cite{mumucolliders}.
Most of the content of this note is indeed addressed to physics
of particular relevance to future muon colliders
(for general reviews see  \cite{physics}). The importance of
analyzing accurately an $s$-channel resonance at a lepton collider and the
physical interest to be assigned to the extracted information are, of course,
not new subjects. Much literature has in fact been devoted to
the problem of the $Z$ resonance shape at electron-positron colliders and to
the implications of its accurate measurement. Many of the special
issues that arise when studying a narrow $s$-channel resonance,
such as a light SM Higgs boson, have also been considered \cite{bbgh}.

However, these latter studies assumed that the beam energy profile
is perfectly known. If a resonance is so narrow that
its width is smaller than or comparable to the beam energy spread,
uncertainty in the beam's energy profile can introduce substantial errors
in the experimental determinations of the resonance parameters.
One of the main goals of the present paper is to assess these errors.
To this end, we shall present a detailed discussion of
methods for studying a narrow $s$-channel resonance or systems of nearly
degenerate resonances at a lepton collider, and consider applications to
particular cases. We carefully analyze the errors
in the measured resonance parameters (such as the total
width and partial widths) that arise from the uncertainty
in the energy spread of the beam. Procedures for reducing this
source of errors are studied and experimental observables with
minimal sensitivity to beam energy spread are emphasized.
Our study will assume that the beam energy spread is independently
measured, as will normally be possible. Expectations and procedures
for this determination at a muon collider will be noted.

Although the parameters of a narrow $s$-channel resonance can be
determined with minimal sensitivity to beam energy
spread by measuring the total Breit-Wigner area, the peak total cross
section and the cross sections in different final state channels,
these measurements are not all easily performed with good
statistical accuracy.  Generally, a more
effective method for determining resonance parameters is
to scan the resonance using specific on- and off-resonance energy
settings. In this paper, we also discuss the
possibly superior new method introduced in Ref.~\cite{ratio}
in which one operates the collider always with center of mass
energy equal to the resonance mass but uses two different beam energy
spreads: one smaller than the resonance width and
one larger. All of these 
possibilities will be compared. The most immediate application
is to the study of a light SM-Higgs, where, as already known
\cite{bbgh,gunmumu,higgsmumu}, accurate
measurements at a muon collider might make it possible
to distinguish a standard model Higgs from the lightest Higgs of a
supersymmetric model. We shall also analyze in detail resonant production
of the lightest pseudogoldstone boson of dynamical electroweak symmetry
breaking models. Following that, we consider two almost coincident
vector resonances of the Degenerate BESS model \cite{DBESS}
when their mass splitting is larger than the beam energy spread.

We shall then examine the case of two nearby  resonances with
mass splitting  much smaller than their average mass, assuming that
the energy spread of the beam is of the same order as the mass splitting.
Different possible procedures for
reducing the errors in the determination of the physical parameters
will be examined. As an
application we shall discuss such a situation in Degenerate BESS.

In Section 2 we present the analysis for a narrow resonance and in Section
3 the applications we have mentioned for this case. Section 4 gives the
analysis for nearly degenerate resonances, and Section 5 a corresponding
application.

\bigskip
\section{Analysis for a narrow resonance}
\label{general}
\bigskip
In this Section we will discuss several ways of determining the
parameters of a narrow resonance at a lepton collider.

For a given decay channel, an $s$-channel resonance can be described by three
parameters: the mass, the total width and the peak cross section.
These quantities can only be accurately measured if the beam parameters
are very well known. Measurement of the mass requires
precise knowledge of the absolute energy value.
Measurement of the width and cross section requires excellent
knowledge of the energy spread of the beam and the ability to determine
the difference between two beam energy settings with great precision.

All of these beam parameters must be known with extraordinary accuracy
in order to study a very narrow resonance. For example, consider
a light SM Higgs boson. In \cite{bbgh} it was found that
the beam energy $E_{\rm beam}$ must be known to better than 1 part in $10^6$
and the beam energy spread $\Delta E_{\rm beam}$ 
must be smaller than 1 part in $10^4$
in order to scan the resonance and determine its width and other
parameters. Below, we shall find that errors 
in these parameters due to uncertainty
in $\Delta E_{\rm beam}$ will only be small compared to 
statistical errors associated with the measurements of a typical cross section
(equivalent to a rate in a particular final state channel) if the error in the
measurement of $\Delta E_{\rm beam}$ is smaller than $\sim 1\%$.

The independent measurement of $\Delta E_{\rm beam}$ can have both
statistical and systematic error.  Let us first consider the 
impact of statistical
errors. We will argue that statistical uncertainties in $\Delta E_{\rm beam}$
(and also $E_{\rm beam}$ itself) should not be significant.
Consider measurement of some resonance parameter $p$ through a series
of observations over a large number ($S$) of spills. In a year of
operation at a muon collider there will be something like $S=10^8$ spills.
Each spill (\ie\ each muon bunch) will contribute
to the measurement of $p$, but the value of $p$ interpreted
from the cross section $\sigma$ observed for a given spill will be uncertain
by an amount $\delta p$. This uncertainty arises (a)
because of the limited number of events $N_i$ accumulated during the spill
and (b) due to uncertainties in 
$E_{\rm beam}$ and $\Delta E_{\rm beam}$ for that spill.
We have, using the short hand notation of $E_b$ for $E_{\rm beam}$,
\begin{equation}
\left[\delta p\over p\right]_i=\left[
{ c^2\over N_i}+ \left(a
{\delta E_b\over E_b}\right)_i^2  +
\left(b
{\delta \Delta E_b\over \Delta E_b}\right)_i^2 
\right]^{1/2}\,,
\label{erroreq}
\end{equation}
where $c={d\ln p\over d \ln \sigma}$,
$a={d\ln p\over d\ln E_b}$ and $b={d\ln p\over d\ln \Delta E_b}$.
$E_b$ and $\Delta E_b$ are measured
by looking at oscillations in the spectra of the muon decay products
arising due to spin precession of the muons in the bunch
during the course of a very large number of turns
around the ring.  These measurements are completely independent
of the measured rate(s) in question. The fractional 
statistical errors ${\delta E_b\over E_b}$
and ${\delta \Delta E_b\over \Delta E_b}$ for each spill are expected
to be small as we shall summarize below. As a result, the contribution
from the ${c^2\over N_i}$ term in Eq.~(\ref{erroreq}) will normally
be much larger than that from the $\delta E_b$ and $\delta\Delta E_b$
terms unless the coefficients $a$ and $b$ are very large. We will
later learn that $a$ and $b$ will always be small under circumstances
in which systematic errors in $E_b$ and $\Delta E_b$ do not
badly distort the parameter measurement. To the extent that this
is true, the 
statistical error $\delta p/p$ after a large number of spills,
computed as
\begin{equation}
{\delta p \over p}=\left[
\sum_i {1\over \left({\delta p\over p}\right)_i^2}\right]^{-1/2}\,,
\label{finalerror}
\end{equation}
will be dominated by the usual $c/\sqrt{\sum_i N_i}$ term. Note that if
this term were altogether absent, then ${\delta p\over p}$ would 
be proportional to $1/\sqrt S$ times
the per spill error from $\delta E_b$ and $\delta \Delta E_b$.
Thus, even if this latter were quite substantial, it would be strongly
suppressed after accumulating data over $10^8$ spills.

Since $E_{\rm beam}$ and $\Delta E_{\rm beam}$ 
are expected to vary somewhat from spill to spill, what is important
is the statistical error with which they can be measured in a given spill.
This has been discussed in 
\cite{spread,raja,reportusa,reporteurope}.
Very roughly, the frequency of
oscillation in the signal of secondary positrons from
the muon decays (which signal is sensitive to the precession
of the naturally-present polarization of the muons)
determines $E_{\rm beam}$ and the decay with time of the oscillation signal
amplitude determines $\Delta E_{\rm beam}/E_{\rm beam}$.
At a resonance mass of $100\gev$,~\footnote{The precise
figures given here are from \cite{reporteurope}.} a beam energy spread
of $\Delta E_{\rm beam}/E_{\rm beam}\sim 3\times 10^{-5}$ can be achieved
by an appropriate machine design
while maintaining adequate yearly luminosity ($\geq 0.1\fbi$).
Further, in this case, {\it for each muon spill} ({\ie\ each
muon bunch}), one can determine the beam
energy itself to 1 part in $10^7$ (5 keV) and measure
the actual magnitude of $\Delta E_{\rm beam}/E_{\rm beam}$ with
an accuracy of roughly $1.67\%$.
For $\Delta E_{\rm beam}/E_{\rm beam}\sim 10^{-3}$, as might
be useful for a much broader resonance than the SM Higgs boson
and as would allow for substantially larger yearly integrated luminosity
($\geq 1\fbi$), $E_{\rm beam}$ can be measured to 2 parts in $10^{6}$ (100 keV)
and $\Delta E_{\rm beam}/E_{\rm beam}$ can be measured with
an accuracy of roughly $0.2\%$. As described above, this level
of statistical error for $E_{\rm beam}$ and $\Delta E_{\rm beam}/E_{\rm beam}$ 
will result in only a tiny $\delta p/p$ error for a given resonance
parameter unless the coefficients $a$ and $b$ of Eq.~(\ref{erroreq})
are very large. Thus, in what follows, we will analyze the parameter
errors introduced by an uncertainty in $\Delta E_{\rm beam}/E_{\rm beam}$ 
that is assumed to be systematic in nature. 
The level of such uncertainty is not well understood at the moment.
The energy spread will be affected by possible time dependence of other
beam parameters, such as emmittance, by time-dependent backgrounds to
the precession measurements, and by other sources of depolarization.
Absent a detailed design for the machine and the polarimeter,
\cite{reporteurope} states that it is ``quite certain'' that the energy
spread can be known with relative systematic error better than $\sim 1\%$,
even if $\Delta E_{\rm beam}/E_{\rm beam}=3\cdot 10^{-5}$.

Given the precision with which the beam energy can
be determined, it is clear that the mass of a resonance
will be precisely known.
We will not consider here the statistical
errors in the measurements of the relevant cross sections,
but these will later become important when choosing among
the different procedures that we will discuss.
Our main focus will be on
the errors in the determination of branching ratios and
total widths induced by a systematic uncertainty in the energy spread of
the beam. Therefore, we will discuss the possibility of
reducing such errors, and also discuss measurements which are
independent of the energy spread. As already noted,
these errors are expected to be important only
when the width of the resonance is smaller than or comparable
to the energy spread of the beam.

We assume that the resonance is well described by a
Breit-Wigner shape. For a resonance $R$ of spin $j$
produced in the $s$-channel and decaying into a
given final state $F$, one has
\begin{equation}\label{BW}
\sigma^{F}(E)=4\pi(2 j+1)\frac{\Gamma(R\to\ell^+\ell^-)
\Gamma(R\to F)}{(E^2-M^2)^2+M^2\Gamma^2}
\end{equation}
where $M$ and $\Gamma$ are the mass and the width of the
resonance, and $E=\sqrt{s}$. We will work in the narrow
width approximation and therefore neglect the running of the
width. We will consider also the total production  cross section
\begin{equation}\label{BWT}
\sigma(E)=4\pi(2 j+1)\frac{\Gamma(R\to\ell^+\ell^-)
\Gamma}{(E^2-M^2)^2+M^2\Gamma^2}
\end{equation}
We assume that in the absence of bremsstrahlung
the lepton beams have a Gaussian beam energy spread specified
by $\Delta E_{\rm beam}/E_{\rm beam}=0.01\,R(\%)$, leading to
a spread in the center of mass energy given by
\begin{equation}\label{spread}
\sigma_E={0.01\,R(\%)\over\sqrt 2}\,E\sim 0.007\,R(\%)\,E\,.
\end{equation}
In the absence of bremsstrahlung, the
energy probability distribution $f(E)$ can be written in the scaling form
\begin{equation}\label{scaling}
f(E)=\frac 1{\sigma_E} g\left(\frac {E-E_0}{\sigma_E}\right)\,,
\end{equation}
where $E_0$ is the central energy setting and
\begin{equation}\label{normalization}
\int f(E)\,dE=1\,.
\end{equation}
Accurate measurements of a narrow resonance require an accurate knowledge
of both $\sigma_E$ and of the shape function $g$.
We will explore the systematic uncertainties associated with errors
in determining $\sigma_E$, assuming that the shape function $g$
is known to be of a Gaussian form.

After including bremsstrahlung (at an $\epem$ collider, beamstrahlung
must also be included) it is no longer possible to write
the full energy distribution in a scaling form; bremsstrahlung has
intrinsic knowledge of the mass of the lepton, which in leading
order enters in the form $\log(E/m_\ell)$.
Since the bremsstrahlung distribution is completely known it may be
convoluted with the scaling form of Eq.~(\ref{scaling})
to obtain a final energy distribution which depends upon both $\sigma_E$
and the (very accurately known) machine energy setting.
Some useful figures illustrating the effects of bremsstrahlung at
a muon collider are Figs. 30, 31 and 32 in Appendix A of \cite{bbgh}.
One typically finds that the peak luminosity at the central beam energy, $E_0$,
is reduced to about 60-80\% of the value in the absence of
bremsstrahlung, and that $d{\cal L}/dE$ falls below
$10^{-3}\left.d{\cal L}/dE\right|_{E=E_0}$ at $E\sim 0.9 E_0$.
We will discuss results at a muon collider
both before and after including bremsstrahlung, and show that
the errors introduced by uncertainty in the beam energy spread
can be very adequately assessed without including bremsstrahlung.

The measured cross sections are obtained from the convolution of
the theoretical cross sections [see (\ref{BW})] with the energy
distribution:
\begin{equation}\label{convolution}
\sigma_c^{F}(E)=4\pi(2j+1) \Gamma(R\to\ell^+\ell^-)
\Gamma(R\to F) h(\Gamma,\sigma_E,E)
\end{equation}
where
\begin{equation}\label{convolution2}
h(\Gamma,\sigma_E,E)=\int \frac{f(E-
E')}{({E'}^2-M^2)^2+M^2\Gamma^2}d E'
\end{equation}
From these equations we can immediately draw  two useful
consequences:
\begin{itemize}
\item  the ratio of $\sigma_c^{F}$ to
the production cross section is given by
\begin{equation}\label{ratio}
\frac{\sigma_c^{F}}{\sigma_c}=B_{F},~~~B_{F}=B(R\to F)
\end{equation}
and does not depend on $\sigma_E$;
\item  due to the
normalization condition (\ref{normalization}), the integral of
$\sigma_c^{F}$ is independent of $\sigma_E$
\begin{equation}\label{integral}
\int\sigma_c^{F}(E)dE=\int\sigma^{F}(E)dE \,.
\end{equation}
Of course, one must be very certain that the integral over $E$ covers
all of the resonance and all of the beam spread, including the
low-energy bremsstrahlung tail.
\end{itemize}
Let us now consider the narrow resonance approximation. In
this case, we can write
\begin{equation}\label{NRA}
M^2\sigma^{F}(E)=\pi(2 j+1)B_{\ell^+\ell^-} B_{F}
\frac{\gamma^2}{x^2+\gamma^2/4}
\end{equation}
where we have scaled the energy and the total width in terms of
the energy spread evaluated at the peak (here we take
$\sigma_E\approx \sigma_M$)
\begin{equation}\label{newvariables}
x=\frac{E-M}{\sigma_M},~~~~\gamma=\frac{\Gamma}{\sigma_M}
\end{equation}
The convolution can also be written in terms of these variables,
with the result (see Eq. (\ref{scaling}))
\begin{eqnarray}\label{scaled-conv}
M^2\sigma_c^{F}(E)&=&\pi(2j+1)B_{\ell^+\ell^-}B_{F}
\int g(x-x')\frac{\gamma^2}{{x'}^2+\gamma^2/4}dx'\nn\\&\equiv
&B_{\ell^+\ell^-}B_{F}\,\Phi(x,\gamma)
\end{eqnarray}
In particular, for the production cross section, we get
\begin{equation}\label{scaled-conv3}
M^2\sigma_c(E)=B_{\ell^+\ell^-}\,\Phi(x,\gamma)
\end{equation}
Some simple results that are valid in the narrow resonance approximation
are the following:
\begin{itemize}
\item If $\Gamma\gg\sigma_M$ (and $\Gamma$ is also large
compared to bremsstrahlung energy spread), then
\begin{equation}
\sigma_c(E)={4\pi B_{\ell^+\ell^-}\over M^2}{\Gamma^2M^2\over
(E^2-M^2)^2+\Gamma^2M^2}\,.
\label{largegam}
\end{equation}
A measurement of the peak cross section (summed over all final state modes)
gives a direct measurement of $B_{\ell^+\ell^-}$. In addition,
the resonance shape can be scanned by using a series
of different energy settings for the machine and $\Gamma$ can
then be determined.
\item If $\Gamma\ll\sigma_M$, then
\begin{equation}
\sigma_c(E)={4\pi B_{\ell^+\ell^-} \over M^2}{\Gamma \sqrt\pi\over 2\sqrt
2\sigma_M}\exp\left[{-(E-M)^2\over 2\sigma_M^2}\right]\,,
\label{smallgam}
\end{equation}
where we have assumed a Gaussian form for the beam energy spread. (We also
neglected bremsstrahlung, but this could be easily included.)
In this case, the peak cross section determines $B_{\ell^+\ell^-}\cdot\Gamma$
{\it provided} $\sigma_M$ is accurately known.
$\sigma_M$ could in turn be measured by changing $E$ relative to $M$.
However, statistical accuracy may be poor because
cross sections are suppressed by $\Gamma/\sigma_M$ in comparison
to those obtained if $\Gamma\geq\sigma_M$.
\end{itemize}
When scanning a resonance,
the above limits make it clear that the cross section is largest
and that systematic errors associated with $\sigma_M$ are smallest
when $\sigma_M$ is as small compared to $\Gamma$ as possible.
However, for a sufficiently narrow resonance, the luminosity
reduction associated with achieving $\sigma_M$
values significantly smaller than $\Gamma$ is often so great
that statistical uncertainties become large.  Thus, we are often
faced with a situation in which $\Gamma$ is comparable
to or only a few times larger than $\sigma_M$.
(For example, this will be the case when studying a light SM Higgs boson
at a muon collider.) It is in this
situation that systematic errors in the resonance
parameters deriving from uncertainty in 
$\sigma_M$ can be enhanced.

We now discuss four ways of extracting $B_{\ell^+\ell^-}$,
$B_{\ell^+\ell^-}B_{F}$ in a specific final state channel
and $\Gamma$ from the data and assess the extent to which their
determination will be influenced by uncertainty associated
with an independent measurement of $\sigma_M$.

\begin{table}[t]
\caption{ Errors on $\Gamma$ and $B$
induced by $\Delta\sigma_M/\sigma_M= +0.05$ (upper lines) and
$\Delta\sigma_M/\sigma_M= -0.05$ (lower lines) evaluated by
choosing one observation at the peak and the other at
$E-M=k\,\Gamma$, for $k=1,2,3$. Bremsstrahlung is neglected.}
\begin{center}
\begin{tabular}{|c|l|l|l|l|l|l|l|l|c|}
\hline
 &
\multicolumn{2}{|c|}{$1\,\Gamma$}&\multicolumn{2}{|c|}{$2\,\Gamma$}
&\multicolumn{2}{|c|}{$3\,\Gamma$}
\\
$\Gamma/\sigma_M$& $\Delta B/B$& $\Delta\Gamma/\Gamma$&$\Delta B/B$&
$\Delta\Gamma/\Gamma$&$\Delta B/B$&$\Delta\Gamma/\Gamma$
\\\hline
 1 & ${+0.220}$ & ${-0.205}$ & ${+0.171}$
 & ${-0.160}$ & ${+0.114}$ & ${-0.101}$\\
& ${-0.140}$ & ${+0.198}$ & ${-0.119}$
 & ${+0.153}$ & ${-0.089}$ & ${+0.094}$\\
\hline
 2 & ${+0.065}$ & ${-0.075}$ & ${+0.043}$
 & ${-0.037}$ & ${+0.035}$ & ${-0.023}$\\
& ${-0.053}$ & ${+0.072}$ & ${-0.039}$
 & ${+0.035}$ & ${-0.033}$ & ${+0.023}$\\
\hline
 3 & ${+0.031}$ & ${-0.036}$ & ${+0.023}$
 & ${-0.017}$ & ${+0.022}$ & ${-0.013}$\\
& ${-0.027}$ & ${+0.034}$ & ${-0.022}$
 & ${+0.017}$ & ${-0.021}$ & ${+0.013}$\\
\hline
 4 & ${+0.018}$ & ${-0.020}$ & ${+0.016}$
 & ${-0.011}$ & ${+0.015}$ & ${-0.009}$\\
& ${-0.017}$ & ${+0.019}$ & ${-0.015}$
 & ${+0.010}$ & ${-0.014}$ & ${+0.009}$\\
\hline
\end{tabular}
\end{center}
\label{nobremi}
\end{table}
\bigskip

\bigskip
\noindent {\bf{Scan of the resonance}} - If $\Gamma\gsim\sigma_M$,
one of the most direct ways to determine the parameters of
an $s$-channel resonance is through a three-point
scan \cite{bbgh}. In this method one measures the production
cross section $\sigma_c$ at three different energies. However,
the position of the peak is independent of the energy spread, and
we do not expect that uncertainty in the latter will induce an
error in the mass of the resonance (we have checked explicitly
that this is indeed the case). Therefore we will assume in the
following that the mass of the resonance is known with high
accuracy, and,  as a consequence,  we will take into account only
two scan points; one will be chosen at the peak and the
other one off the peak. We will shortly discuss how the
results depend on the position of this last point. The parameters
$(B,\Gamma)$ are then extracted by a two parameter
fit. Here, $B=B_{\ell^+\ell^-}$ if one is able to sum
over all final state modes, and $B=B_{\ell^+\ell^-}B_{F}$
if one focuses on a particular final state.
In principle this problem has a unique solution, by
deconvoluting the observed cross section. However, the error in
the determination of $\sigma_M$ (the energy spread at the peak of
the resonance) induces an error on the determination of the
parameters of the resonance. Assuming that the measured value of
$\sigma_c$ at a given energy $E$ is given by
$\sigma_c(E,B,\Gamma,\sigma_M)$ (here again we use the narrow
resonance approximation to put $\sigma_E\approx\sigma_M$), the
changes in the values of $B$ and $\Gamma$ that result if $\sigma_M$
is shifted by an amount $\Delta\sigma_M$
can be evaluated through the equation
\begin{equation}\label{condition1}
\sigma_c(E,B,\Gamma,\sigma_M)=
\sigma_c(E,B+\Delta B,\Gamma+\Delta\Gamma,\sigma_M+\Delta\sigma_M)
\end{equation}
or, from Eq. (\ref{scaled-conv3}),
\begin{equation}\label{condition2}
\Phi(x,\gamma)=
\Phi(x+\Delta x,\gamma+\Delta\gamma)
\left(1+\frac{\Delta B}{B}\right)
\end{equation}
where
\begin{eqnarray}\label{def}
\Delta x&=&x(\sigma_M+\Delta\sigma_M)-x(\sigma_M)\approx
-x\frac{\Delta\sigma_M}
{\sigma_M}\nn\\\Delta\gamma&=&\gamma(\sigma_M+\Delta\sigma_M)-\gamma(\sigma_M)
\approx \frac{\Gamma}{\sigma_M}\left(
\frac{\Delta\Gamma}{\Gamma}-\frac{\Delta\sigma_M}{\sigma_M}\right)
\end{eqnarray}
If we measure the cross section at two different energies, and assume that 
$\Delta\sigma_M$ is the same at these two nearby energies
(as would be typical of a systematic error), we can easily
employ Eq. (\ref{condition2}) to determine the induced
fractional shifts $\Delta\Gamma/\Gamma$ and $\Delta B/B$. We fix one of the two
points at the peak ($x=0$), and we let the other one vary
between $E-M=\Gamma$ and $E-M=3\Gamma$. The induced
systematic errors are illustrated in
a series of Tables. In Tables \ref{nobremi} and \ref{nobremii},
we assume a Gaussian energy distribution without including bremsstrahlung
and tabulate errors induced by
$\Delta\sigma_M/\sigma_M=0.05$ and 0.01, respectively. As clear
from Eq. (\ref{condition2}), the fractional errors on $\Gamma$ and
$B$ depend only on the $E-M$ choice and on the reduced width $\gamma$,
but not on $B$.

\begin{table}[h]
\caption{Same as Table \protect\ref{nobremi} for
$\Delta\sigma_M/\sigma_M= +0.01$ (upper lines) and
$\Delta\sigma_M/\sigma_M= -0.01$ (lower lines).}
\begin{center}
\begin{tabular}{|c|l|l|l|l|l|l|l|l|c|}
\hline
 &
\multicolumn{2}{|c|}{$1\,\Gamma$}&\multicolumn{2}{|c|}{$2\,\Gamma$}
&\multicolumn{2}{|c|}{$3\,\Gamma$}
\\
$\Gamma/\sigma_M$& $\Delta B/B$& $\Delta\Gamma/\Gamma$&$\Delta B/B$&
$\Delta\Gamma/\Gamma$&$\Delta B/B$&$\Delta\Gamma/\Gamma$
\\\hline
 1 & ${+0.036}$ & ${-0.040}$ & ${+0.029}$
 & ${-0.031}$ & ${+0.020}$ & ${-0.020}$\\
& ${-0.033}$ & ${+0.040}$ & ${-0.027}$
 & ${+0.031}$ & ${-0.019}$ & ${+0.019}$\\
\hline
 2 & ${+0.012}$ & ${-0.015}$ & ${+0.008}$
 & ${-0.007}$ & ${+0.007}$ & ${-0.004}$\\
& ${-0.011}$ & ${+0.015}$ & ${-0.008}$
 & ${+0.007}$ & ${-0.007}$ & ${+0.004}$\\
\hline
 3 & ${+0.006}$ & ${-0.007}$ & ${+0.005}$
 & ${-0.003}$ & ${+0.004}$ & ${-0.003}$\\
& ${-0.006}$ & ${+0.007}$ & ${-0.004}$
 & ${+0.003}$ & ${-0.004}$ & ${+0.003}$\\
\hline
 4 & ${+0.004}$ & ${-0.004}$ & ${+0.003}$
 & ${-0.002}$ & ${+0.003}$ & ${-0.002}$\\
& ${-0.003}$ & ${+0.004}$ & ${-0.003}$
 & ${+0.002}$ & ${-0.003}$ & ${+0.002}$\\
\hline
\end{tabular}
\end{center}
\label{nobremii}
\end{table}

\begin{figure}[p]
\epsfysize=16truecm
\centerline{\epsffile{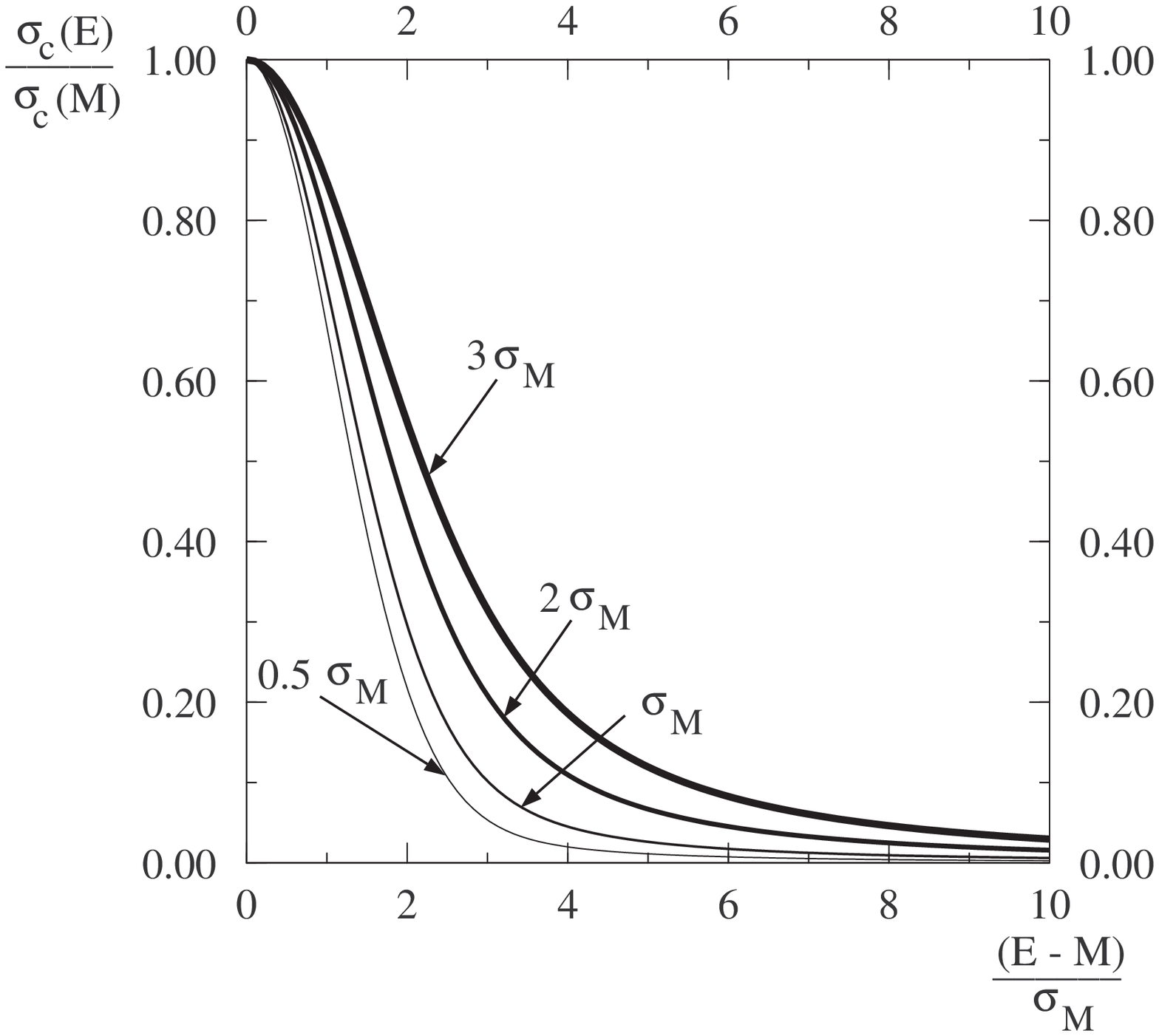}}
\noindent
\caption{
Plot of the ratio $\sigma_c(E)/\sigma_c(M)$ vs.
$x=(E-M)/\sigma_M$, for  $\Gamma=k\sigma_M$, with $k=0.5,1,2,3$.}
\label{fig1}
\end{figure}

As expected from our earlier discussion, the fractional errors become
smaller for larger $\Gamma/\sigma_M$.
The fractional errors are also decreased by moving the
second point of the scan away from the peak (see Tables \ref{nobremi}
and \ref{nobremii}).
On the other hand the statistical errors for $\sigma_c$ are increasing as
\begin{equation}\label{ratio2}
\sigma_c(E)/\sigma_c(M) =\frac{\Phi(x,\gamma)}{\Phi(0,\gamma)}
\end{equation}
becomes smaller. (Due to presence of background, the decrease
is not simply proportional to $\sqrt{\sigma_c(M)/\sigma_c(E)}$.)
By fixing the amount of luminosity that one can safely lose
without incurring substantial statistical error in $\sigma_c(E)$, and for
a given ratio $\Gamma/\sigma_M$, one can, by looking at Fig. \ref{fig1},
fix the second scan point. For instance, for $\Gamma=2\sigma_M$,
allowing for a loss in luminosity of 60\%, we see that the point to
be chosen is $x\approx 2$ corresponding to $E\approx M+\Gamma$.
By choosing the second point of the scan at $E=M+2\Gamma$, the
resulting behaviour of the systematic errors vs. $\gamma$ is illustrated in
Fig. \ref{fig2}. For a given uncertainty on the energy spread the errors
decrease for increasing $\gamma$. Therefore, as noted earlier,
one should employ
the smallest value of $\sigma_M$ consistent with having luminosity
large enough to give small statistical $\sigma_c$ errors. For
instance, at $\gamma=1$ the errors on the resonance parameters
are of order $2.5\div 3.5$ times $\Delta\sigma_M/\sigma_M$,
whereas for $\gamma=3$ they are down to $0.3\div 0.5$ times
$\Delta\sigma_M/\sigma_M$.

Of course, a full study of the
optimization of this procedure in a concrete case requires also
taking into account the third scan point. In fact, one has
the  further freedom of moving the position of this point,
with corresponding changes in the errors. One can show that by taking
the third scan point in a symmetric position with respect to the
other off-peak point, one gets the same errors shown in
Tables \ref{nobremi} and  \ref{nobremii} (in the absence of bremsstrahlung).
By taking an asymmetric configuration the results are
in between the ones obtained with the corresponding
symmetric configurations. For instance, by taking one point at
the peak, one at $E=M+2\Gamma$ and the third one at $E=M-\Gamma$,
we find, for $\gamma=1$ and $\Delta\sigma_M/\sigma_M=0.05$,
$\Delta\Gamma/\Gamma=-0.174$ and $\Delta B/B=+0.186$.

\begin{table}[t]
\caption{Errors for $\Gamma$ and $B$
induced by $\Delta\sigma_M/\sigma_M= +0.05$ for no bremsstrahlung
(first lines, from Table \ref{nobremi}), and after including bremsstrahlung
for $R=0.003\%$ (second lines),
$R=0.03\%$ (third lines) and $R=0.1\%$ (fourth lines).
The energy profile before bremsstrahlung is taken to be
a Gaussian specified by the $\sigma_M$
corresponding to a given $R$ taking $M=100\gev$.}
\begin{center}
\begin{tabular}{|c|l|l|l|l|l|l|l|l|c|}
\hline
 &
\multicolumn{2}{|c|}{$1\,\Gamma$}&\multicolumn{2}{|c|}{$2\,\Gamma$}
&\multicolumn{2}{|c|}{$3\,\Gamma$}
\\
$\Gamma/\sigma_M$& $\Delta B/B$& $\Delta\Gamma/\Gamma$&$\Delta B/B$&
$\Delta\Gamma/\Gamma$&$\Delta B/B$&$\Delta\Gamma/\Gamma$
\\\hline
1 & ${+0.22}$ & ${-0.21}$ & ${+0.17}$ & ${-0.16}$ & ${+0.11}$ & ${-0.10}$\\
 & ${+0.22}$ & ${-0.21}$ & ${+0.18}$ & ${-0.16}$ & ${+0.12}$ & ${-0.11}$\\
 & ${+0.22}$ & ${-0.21}$ & ${+0.18}$ & ${-0.16}$ & ${+0.12}$ & ${-0.11}$\\
 & ${+0.22}$ & ${-0.21}$ & ${+0.18}$ & ${-0.16}$ & ${+0.12}$ & ${-0.11}$\\
\hline
2 & ${+0.065}$ & ${-0.075}$ & ${+0.043}$ & ${-0.037}$ & ${+0.035}$ & ${-0.023}$\\
& ${+0.067}$ & ${-0.078}$ & ${+0.046}$ & ${-0.044}$ & ${+0.040}$ & ${-0.032}$\\
& ${+0.065}$ & ${-0.075}$ & ${+0.044}$ & ${-0.041}$ & ${+0.037}$ & ${-0.029}$\\
& ${+0.067}$ & ${-0.077}$ & ${+0.045}$ & ${-0.042}$ & ${+0.039}$ & ${-0.031}$\\
\hline
3 & ${+0.031}$ & ${-0.036}$ & ${+0.023}$ & ${-0.017}$ & ${+0.022}$ & ${-0.013}$\\
& ${+0.031}$ & ${-0.035}$ & ${+0.024}$ & ${-0.020}$ & ${+0.023}$ & ${-0.017}$\\
& ${+0.031}$ & ${-0.037}$ & ${+0.024}$ & ${-0.020}$ & ${+0.023}$ & ${-0.017}$\\
& ${+0.031}$ & ${-0.038}$ & ${+0.024}$ & ${-0.021}$ & ${+0.022}$ & ${-0.016}$\\
\hline
4 & ${+0.018}$ & ${-0.020}$ & ${+0.016}$ & ${-0.011}$ & ${+0.015}$ & ${-0.009}$\\
& ${+0.019}$ & ${-0.022}$ & ${+0.017}$ & ${-0.013}$ & ${+0.015}$ & ${-0.011}$\\
& ${+0.018}$ & ${-0.021}$ & ${+0.016}$ & ${-0.012}$ & ${+0.015}$ & ${-0.011}$\\
& ${+0.018}$ & ${-0.020}$ & ${+0.015}$ & ${-0.012}$ & ${+0.015}$ & ${-0.011}$\\
\hline
\end{tabular}
\end{center}
\label{brema}
\end{table}
\begin{table}[h]
\caption{Errors for $\Gamma$ and $B$
expressed as $\Delta \Gamma/\Gamma=c_\Gamma \Delta\sigma_M/\sigma_M$
and $\Delta B/B=c_B \Delta\sigma_M/\sigma_M$ in the limit of
small $\Delta\sigma_M/\sigma_M$. Same entry ordering and bremsstrahlung
assumptions as in Table~\ref{brema}.}
\begin{center}
\begin{tabular}{|c|l|l|l|l|l|l|l|l|c|}
\hline
 &
\multicolumn{2}{|c|}{$1\,\Gamma$}&\multicolumn{2}{|c|}{$2\,\Gamma$}
&\multicolumn{2}{|c|}{$3\,\Gamma$}
\\
$\Gamma/\sigma_M$ & $~~~c_B$ & $~~~c_\Gamma$ & $~~~c_B$ & $~~~c_\Gamma$ &
$~~~c_B$ & $~~~c_\Gamma$
\\\hline
 1 & 3.4 & $-$4.0 & 2.8 & $-$3.1 & 2.0 & $-$2.0 \\
   & 3.6 & $-$4.3 & 2.8 & $-$3.2 & 2.1 & $-$2.1 \\
   & 3.2 & $-$3.7 & 2.8 & $-$3.2 & 2.1 & $-$2.1 \\
   & 3.4 & $-$3.9 & 2.8 & $-$3.2 & 2.1 & $-$2.2 \\
\hline
 2 & 1.2 & $-$1.5 & 0.81 & $-$0.72 & 0.69 & $-$0.45 \\
   & 1.2 & $-$1.5 & 0.85 & $-$0.81 & 0.72 & $-$0.59 \\
   & 1.2 & $-$1.5 & 0.86 & $-$0.82 & 0.74 & $-$0.61 \\
   & 1.2 & $-$1.5 & 0.85 & $-$0.82 & 0.74 & $-$0.61 \\
\hline
 3 & 0.58 & $-$0.70 & 0.46 & $-$0.34 & 0.43 & $-$0.26 \\
   & 0.62 & $-$0.76 & 0.48 & $-$0.40 & 0.43 & $-$0.32 \\
   & 0.60 & $-$0.76 & 0.48 & $-$0.42 & 0.46 & $-$0.36 \\
   & 0.57 & $-$0.71 & 0.48 & $-$0.42 & 0.44 & $-$0.36 \\
\hline
 4 & 0.35 & $-$0.40 & 0.31 & $-$0.21 & 0.30 & $-$0.17 \\
   & 0.39 & $-$0.45 & 0.30 & $-$0.24 & 0.30 & $-$0.26 \\
   & 0.39 & $-$0.43 & 0.32 & $-$0.23 & 0.29 & $-$0.23 \\
   & 0.35 & $-$0.40 & 0.29 & $-$0.24 & 0.28 & $-$0.20 \\
\hline
\end{tabular}
\end{center}
\label{derapprox}
\end{table}

Inclusion of bremsstrahlung might complicate the picture
developed above, since the induced errors could, in principle,
depend upon the resonance mass and the value
of $R$, even when holding $\Gamma/\sigma_M$ and $(E-M)/\sigma_M$ fixed.
To determine if the effects of bremsstrahlung are important
for determining systematic errors coming from uncertainty in $\sigma_M$,
we begin by considering again $\Delta \sigma_M/\sigma_M=+0.05$.
We adopt $M=100\gev$ and consider $R=0.003\%$, $0.03\%$, and $0.1\%$
at a muon collider. The resulting
errors are given in Table~\ref{brema} in comparison to
the no-bremsstrahlung results of the upper lines of Table~\ref{nobremi}.
The induced errors computed including bremsstrahlung are quite independent
of $R$ and (within the numerical errors of our programs) are
essentially the same as those computed in the absence
of bremsstrahlung. This can be further illustrated in the limit
of very small $\Delta\sigma_M/\sigma_M$.  In this limit, one can simply
compute the errors induced in $B$ and $\Gamma$ by using a linear
expansion of Eq.~(\ref{condition1}) and solving the resulting matrix equation.
The results for $\Delta B/B$ and $\Delta \Gamma/\Gamma$ expressed
as a coefficient times $\Delta\sigma_M/\sigma_M$ are given in Table
\ref{derapprox} for four cases. The first case is that of no bremsstrahlung.
(Note that for $\Delta\sigma_M/\sigma_M=\pm 0.01$ the errors predicted
are quite close to those of Table \ref{nobremii}.) The other
three cases are for $M=100\gev$ and $R=0.003\%$, $0.03\%$, and $0.1\%$, 
including bremsstrahlung. Once again, we see that the error coefficients
computed including bremsstrahlung do not depend much on $R$
and are essentially the same (within the numerical errors of our programs)
as the error coefficients computed without
including bremsstrahlung.  Thus, in what follows we will
discuss errors from $\Delta\sigma_M/\sigma_M$ without including
the effects of bremsstrahlung. Once a particular resonance is
discovered, the effects of bremsstrahlung upon the results
presented here can be easily incorporated to whatever precision is required.

As regards $\epem$ colliders, at which bremsstrahlung is a more substantial
effect, a study (similar to that above
for the $\mupmum$ collider) shows that the Gaussian approximation
for studying systematic errors from $\Delta\sigma_M$ is
again reasonable.  Although beamstrahlung is also important
at an $\epem$ collider, we did not study its effects. However,
current $\epem$ collider designs are such that the energy
spreading from beamstrahlung is typically comparable to or smaller than
that from bremsstrahlung.

An interesting general question regarding the scan procedure is how
to optimize the choice of $\sigma_M$ relative to $\Gamma$
so as to minimize the net statistical plus systematic error.
Let us consider using $k=1$ for the scan, assuming a resonance
mass of $M=100\gev$ and
that $\Delta\sigma_M/\sigma_M$ is small ($\lsim 0.02$),
so that the linear error expansion using coefficients $c_{\Gamma}$
and $c_B$ is valid. 
A very rough parameterization for the induced error in $\Gamma$
when $0.2\lsim \Gamma/\sigma_M\lsim 3$ and $\Delta\sigma_M/\sigma_M\lsim 0.02$
is
\begin{equation}
{\Delta\Gamma\over\Gamma}\sim3.3\left({\sigma_M\over\Gamma}\right)^{1.288}
{\Delta\sigma_M\over\sigma_M}\,.
\label{dgamform}
\end{equation}
The optimal choice of $R$ 
for determining the parameters of a given resonance depends
upon many factors: how the machine luminosity varies with $R$; 
the variation of $\Delta\sigma_M/\sigma_M$ with $R$; the 
variation of the $\Delta\sigma_M$-induced
errors as a function of $R$; the magnitude of the resonance width
(in particular, as compared to $\sigma_M$);
and the size of backgrounds in the important final states to
which the resonance decays. At a muon collider, $R$ in the range 
$0.1\%\div0.15\%$ is the natural result and
allows maximal luminosity. Increasing $R$ above this range
does not result in significant luminosity increase; in contrast,
the luminosity declines rapidly as $R$ is decreased below this range.
A convenient parameterization for the luminosity of an $E=100\gev$
muon collider, valid for $0.003\%\lsim R\lsim 0.12\%$, is:~\footnote{The
$L_{\rm year}$ parameterization interpolates the three
results of Table 5 in \cite{reporteurope} taken from \cite{reportusa}.} 
\begin{equation}
L_{\rm year}=1.2\fbi\,\left({R\over 0.12\%}\right)^{0.67362}\,.
\label{lform}
\end{equation}
The specific coefficient in Eq.~(\ref{lform}) represents the most
pessimistic estimate for the instantaneous luminosities.
On occasion we shall also discuss results assuming a factor of 10 larger
coefficient, which we shall refer to as the optimistic luminosity estimate. 
We do not know what to expect for the variation of the systematic
error in $\sigma_M$ as a function of $R$. In order to understand
how the optimization might work, we have adopted on a purely adhoc basis
a form that mimics the per-spill {\it statistical} error:~\footnote{This
form for the statistical
$\Delta\sigma_M/\sigma_M$ error interpolates the results for
the two $R=0.003\%$ and $R=0.1\%$ cases given in Sec. 4.3.2
of \cite{reporteurope} using a power law form.}
\begin{equation}
{\Delta\sigma_M\over\sigma_M}=2\cdot 10^{-3}\,\left({R\over
0.12\%}\right)^{-0.57477}\,.
\label{dsigmaform}
\end{equation}
This form corresponds to a decreasing fractional systematic error
in $\sigma_M$ as $R$ increases, as seems a likely possibility.
Finally, it is useful to recall the value of $\sigma_M$ as a function of $R$:
$$\sigma_M=86\mev\,{M\over 100\gev}\,\left({R\over 0.12\%}\right)\,.$$
For $R=0.003\%$, the smallest $R$ that is likely
to be achievable, these quantities have the values $L_{\rm year}=0.1\fbi$,
${\Delta\sigma_M/ \sigma_M}=0.0167$, and $\sigma_M\sim 2\mev$ (for
$M=100\gev$). The rough implications of these dependencies are as follows.
\begin{itemize}
\item 
For a broad resonance, defined as one
with $\Gamma\gg0.001M$, one should operate the muon collider
at its natural $R$ value of order $0.12\%$. The $\Delta\sigma_M$-induced
errors will be very tiny, both because $\Gamma/\sigma_M$ is very large
and because $\Delta\sigma_M/\sigma_M$ should be small for such $R$.
A precision scan of the resonance is readily possible in this case.
Further, the resonance cross section is maximal for large $\Gamma/\sigma_M$
and both background and signal rates vary slowly as a function of $R$.
\item
For a resonance with $\Gamma<0.001M$, one will wish to reduce $R$
until $\Gamma\gsim \sigma_M$. The primary reasons to avoid
$\Gamma<\sigma_M$ are to avoid the large $\Delta\sigma_M$-induced errors
summarized above and to maintain adequate sensitivity
to the resonance width. Keeping $\Gamma/\sigma_M>1$, one can 
combine the above formulae with the predicted variation
of statistical error as a function of $R$ at fixed $L$ to
determine the optimal $R$ choice. To illustrate, we will focus on the
statistical error for $\Gamma$ obtained using the three-point
scan with measurements at $E=M,M\pm 2\sigm$. If $\Gamma/\sigma_M$ is
large (typically $>7\div 10$) the signal cross sections at the
three scan points are roughly independent of $R$ (but only for $R$ values
smaller than or not too much larger than 
that corresponding to $\sigm$). Since
the background cross section is also independent
of $R$, statistical errors would not vary much with $R$
if the total integrated luminosity is held fixed.
More generally, one will find that, at fixed $L$,
$\Delta\Gamma/\Gamma\propto R^p$ with $p>0$. (This is a very
crude representation valid only for a limited range of $R$.)
The exact rate of increase depends upon both $\Gamma/\sigm$
and the background level. Such cases will be discussed shortly.
However, even if $p=0$, $\Delta\Gamma/\Gamma$ also varies
as $1/\sqrt L$, where the available $L$ is given in Eq.~(\ref{lform}).
Defining $f\equiv R/0.003\%$, the net result is that 
we can write $(\Delta\Gamma/\Gamma)_{\rm stat.}=c_{\rm stat.} f^{p-0.337}$.
Meanwhile, if we insert the expressions for $\sigma_M$
and $\Delta\sigma_M/\sigma_M$ in the expression for 
the $\Delta\sigma_M$-induced $\Delta\Gamma/\Gamma$, we find
$(\Delta\Gamma/\Gamma)_{\rm syst.}=c_{\rm syst.} f^{0.713}$, where
(for the $k=1$ three-point scan) 
$c_{\rm syst.}\sim 0.06\left({2\mev\over\Gamma}\right)^{1.288}$
if $M=100\gev$; we have normalized $\Gamma$ to the approximate
width of a $100\gev$ SM Higgs boson.
If $p-0.337<0$, as would apply for $p\sim 0$, then,
if the statistical and systematic errors are added in quadrature,
the opposite signs of their $f$ exponents means that 
there will be a minimum in the combined error as a function of $f$.
If we use as a reference the value $f_0$ 
such that $c_{\rm stat.}=c_{\rm syst.}$,
then one finds that the optimal $f$ is such that $f/f_0=R/R_0\sim 0.7$
when $p\sim 0$ (requiring $\Gamma/\sigm(R_0)\gg 1$).
In other words, one should choose a value of $R$ somewhat
smaller than that value which would yield equal statistical and systematic 
errors. 

However, large $\Gamma/\sigm$ cannot be achieved for
the very narrow Higgs and pseudogoldstone bosons discussed
later in the paper. Even
for the larger $\mh$ and $\mpzero$ values of possible interest for this
optimization discussion, we have, at best, 
$\Gamma/\sigm(R=0.003\%)\sim 3\div 5$. 
Starting from $R=0.003\%$, one finds that, at fixed total $L$,
$\Delta\Gamma/\Gamma$ increases quite rapidly with increasing $R$.
As two examples, $\Gamma/\sigm(R=0.003\%)\sim 3$ ($\sim 5$)
for a SM-like Higgs boson with $\mh\sim 140\gev$ ($\sim 150\gev$). 
In these two cases,
$(\Delta\Gamma/\Gamma)_{\rm stat.}$ increases by a factor $\sim \sqrt 10$ 
($\sim \sqrt 2$)
as $R$ increases from $0.003\%$ to $0.01\%$.  Representing this
increase using a power law (it is actually somewhat faster than 
a power law), we can roughly write 
$(\Delta\Gamma/\Gamma)_{\rm stat.}\sim c R^p$
at fixed $L$, with $p\sim 1$ ($p\sim 0.3$) for $\mh=140\gev$ 
($\mh\sim 150\gev$). (Note that for $R$ above the $0.003\%$ to $0.01\%$
range, the $p$ values are much bigger.)
From the analysis given in the preceding paragraph,
we immediately see that if $p>0.337$, the best 
statistical and systematic errors will both be achieved by
taking $R$ as small as possible. In the SM Higgs case, $\Gamma/\sigm>5$,
corresponding to $\mh>150\gev$ would be required before 
$p<0.337$ and there could
be some possible gain from increasing $R$. Similarly, in the case of the
narrow pseudogoldstone boson $\pzero$
it is only at the very largest $\mpzero=200\gev$ mass considered
that $p$ falls just slightly below $0.337$. 
In practice, our a priori knowledge of $\Gamma$ 
(from the initial scan needed to pin
down the precise location of the $H$ or $P^0$) will be too imprecise
to allow for such optimization; one should plan on operating
the machine at $R=0.003\%$ if initial information indicates
that we are dealing with a Higgs or pseudogoldstone boson.

\end{itemize}

\bigskip
\noindent {\bf{Measuring the on-peak cross sections for 
different \boldmath$\sigma_M$}} - 
In this technique \cite{ratio} one presumes that $M$ is already quite well
known and that one has in hand a rough idea of the size of $\Gamma$.
(This is the likely result after the first rough scan used to locate
the resonance.) One then operates the collider at $E=M$
for two different values of $\sigma_M$ (spending perhaps a year or two
at each value). The results of Eqs.~(\ref{largegam}) and (\ref{smallgam})
show that if $\sigma_M^{\rm min}\ll\Gamma$ and $\sigma_M^{\rm max}\gg\Gamma$,
then $\sigma_c(\sigma_M^{\rm min})/\sigma_c(\sigma_M^{\rm max})={2\sqrt
2\sigma_M^{\rm max}\over \Gamma\sqrt\pi}$. 
The systematic error in $\Gamma$ is
then given by 
$(\Delta\Gamma/\Gamma)_{\rm syst.}
=\Delta\sigma_M^{\rm max}/ \sigma_M^{\rm max}$.
In practice, $\sigma_M^{\rm max}/\sigma_M^{\rm min}$ 
will be limited in size. If we define 
$\sigma_M^{\rm central}=\sqrt{\sigma_M^{\rm max}\sigma_M^{\rm min}}$
(the geometric mean value) and
compute $r_c\equiv\sigma_c(\sigma_M^{\rm min})/\sigma_c(\sigma_M^{\rm max})=
\Phi(0,\gamma_{\rm min})/\Phi(0,\gamma_{\rm max})$
(where $\gamma_{\rm max,min}=\Gamma/\sigma_M^{\rm max,min}$) as
a function of $\Gamma$, our ability to measure $\Gamma$
in this way for any given value of 
$\sigma_M^{\rm max}/\sigma_M^{\rm min}$ is determined
by the slope $|s|$ of $\ln[\Gamma/\sigma_M^{\rm central}]$ 
plotted as a function of $\ln[r_c]$.  
The relevant plot from Ref.~\cite{ratio} appears in Fig.~\ref{dlgdlr}.
For a known $\sigma_M^{\rm central}$, the $|s|$
at any $\Gamma/\sigma_M^{\rm central}$ gives the relation
$(\Delta\Gamma/\Gamma)_{\rm stat.}=|s|(\Delta r_c/r_c)_{\rm stat.}$,
where $\Delta r_c/r_c$ is computed by combining the
fractional statistical errors for $\sigma_c(\sigma_M^{\rm min})$
and $\sigma_c(\sigma_M^{\rm max})$ in quadrature.  
The point at which the magnitude of the slope, $|s|$, 
is smallest indicates the point
at which a given fractional statistical error in the cross section ratio 
will give the most accurate determination 
(as measured by fractional error) of $\Gamma/\sigma_M^{\rm central}$.
We observe that $\Gamma/\sigma_M^{\rm central}\sim 2\div 3$ gives the
smallest $|s|$ (and hence smallest statistical error),
although $|s|$ at $\Gamma/\sigma_M^{\rm central}\sim 1$ 
is not that much larger. As expected,
the larger $\sigma_M^{\rm max}/\sigma_M^{\rm min}$, the smaller $|s|$
at any given $\Gamma/\sigma_M^{\rm central}$. 
The systematic error in $\Gamma$ due to uncertainties in $\sigma_M^{\rm max}$
and $\sigma_M^{\rm min}$ is obtained from
\begin{equation}
{\Phi\left(0,{\Gamma\over \sigma_M^{\rm min}}\right)\over 
\Phi\left(0,{\Gamma\over\sigma_M^{\rm max}}\right)}=
{\Phi\left(0,{\Gamma+\Delta\Gamma\over \sigma_M^{\rm
min}+\Delta\sigma_M^{\rm min}}\right)\over 
\Phi\left(0,{\Gamma+\Delta\Gamma \over\sigma_M^{\rm max}
+\Delta\sigma_M^{\rm max}}\right)}\,.
\label{dgamratio}
\end{equation}
In most instances, the $\sigma_M^{\rm min}$ measurement is not
very sensitive to the precise value of $\sigma_M^{\rm min}$, in which case
the result is $(\Delta\Gamma/\Gamma)_{\rm syst.}
=\Delta\sigma_M^{\rm max}/\sigma_M^{\rm max}$. 
More generally $(\Delta\Gamma/\Gamma)_{\rm syst.}=\Delta\sigma_M/\sigma_M$
to the extent that $\Delta\sigma_M/\sigma_M$ is the same at the different
$\sigma_M$ settings (as might apply for systematic errors of a certain type).
From this we see one clear advantage of this technique: the systematic
error in $\Gamma$ due to systematic error in $\sigma_M$ does not 
increase with decreasing $\Gamma/\sigma_M^{\rm max,min}$.  If statistical
errors for measuring the two $\sigma_c$'s can be kept small,
this is a clear advantage of the technique as compared to the scan
technique in which small $\Gamma/\sigma_M$ leads to large
$\Delta\sigma_M$-induced systematic error even if statistics
are excellent for all measurements. We will shortly discuss
issues related to statistical errors.

To continue our analysis of systematic errors,
let us note that, in the limit of very high statistics
for the cross section measurements (\ie\ zero statistical error
for $r_c$), precise values of both 
$\gam_{\rm min}=\Gamma/\sigma_M^{\rm min}$ and
$\gam_{\rm max}=\Gamma/\sigma_M^{\rm max}$ are obtained by the above procedure.
Further, $\sigma_c(\sigma_M^{\rm min})$ and $\sigma_c(\sigma_M^{\rm max})$
are given by $B\Phi(0,\gam_{\rm min})$ and $B\Phi(0,\gam_{\rm max})$,
respectively, where $B$ stands for  
$B_{\ell^+\ell^-}$ or $B_{\ell^+\ell^-}B_F$, depending upon whether we are
looking at the total rate or the rate in some particular final channel $F$.
As a result, there is no systematic error in $B$
from uncertainty in $\sigma_M$, only statistical error
associated with the number of events observed. This is another advantage
of this technique.

Of course small systematic errors are not important
if statistical errors for the technique are not also small.
We summarize the considerations \cite{ratio}.
For the SM Higgs and the $\pzero$ we found that it is best to  use
$\sigmmin$ corresponding to $R=0.003\%$ and 
$\sigmmax$ corresponding to $R=0.03\%$, so that 
$\sigm^{\rm central}$ corresponds to $R=0.003\%\times\sqrt{10}$.
The parameterization for the variation of $\lyear$ given
in Eq.~(\ref{lform}) implies that $\lyear=0.1\fbi$ ($0.47\fbi$)
for $R=0.003\%$ ($0.03\%$). If, for example, $\Gamma/\sigm^{\rm central}=1$, 
one finds $\sigma_c(\sigmmin)/\sigma_c(\sigmmax)=4.5$, implying that
the signal rate $S(\sigm)=\lyear(\sigm)\sigma_c(\sigm)$ is nearly 
the same for $\sigmmax$ as for $\sigmmin$. However,
the background rate $B$ is proportional
to $L$ and thus $B/S$ is a factor of 4.7 times larger at $\sigmmax$
than at $\sigmmin$. Consequently, the statistical error in the measurement
of $\sigma_c(\sigmmax)$ will be worse than for 
$\sigma_c(\sigmmin)$ for the same $S$.~\footnote{In the scan procedure,
there is a similar difficulty. There, $B/S$ is large
for the off-peak measurements.} 
For a given running time at a given $\sigm$,
one must compute the channel-by-channel $S$ and $B$ rates,
compute the fractional error in $\sigma_c(\sigm)$
for each channel, and then combine all channels
to get the net $\sigma_c(\sigm)$ error. 
This must be done for $\sigm=\sigmmin$ 
and $\sigm=\sigmmax$. One then computes the net 
$r_c$ and net $\sigma_c$ errors as:
\bea
{\Delta r_c\over r_c}&=&
\left\{\left[{\Delta \sigma_c(\sigmmin)\over\sigma_c(\sigmmin)}\right]^2
+\left[{\Delta \sigma_c(\sigmmax)\over\sigma_c(\sigmmax)}\right]^2
\right\}^{1/2}\,;
\label{rcerror}\cr
{\Delta\sigma_c\over\sigma_c}&=&
\left\{\left[{\Delta \sigma_c(\sigmmin)\over \sigma_c(\sigmmin)}\right]^{-2}
+\left[{\Delta \sigma_c(\sigmmax)\over\sigma_c(\sigmmax)}\right]^{-2}
\right\}^{-1/2}\,.
\label{sigcerror}
\eea
The ratio of running times at $\sigmmin$ vs.
$\sigmmax$ cannot be chosen so as to simultaneously
minimize the net $\Delta\sigma_c/\sigma_c$
and $\Delta r_c/r_c$. The former is minimized by running
only at $\sigmmin$, while the latter is typically
minimized for $t(\sigmmin)/t(\sigmmax)\lsim 1$. For the SM Higgs, 
a good compromise is to take $t(\sigmmin)/t(\sigmmax)=1$. 
As demonstrated in the next section, 
it turns out that for both the $P^0$ and the SM Higgs boson,
the ratio $\lyear(R=0.03\%)/\lyear(R=0.003\%)\sim 4.7$
and the predicted cross sections and backgrounds are such
that this technique is very competitive with the scan
technique as regards statistical errors for $\Gamma$.

Of course, for resonances (such as those of the Degenerate BESS model
considered later) that have fairly large widths, the normal scan
procedure can achieve superior results to the $r_c$-ratio technique.
This is because there will be little sacrifice in luminosity associated
with choosing an $R$ such that $\Gamma/\sigma_M$ is substantially
larger than 1. Measurements on the wings of the resonance will have
nearly the same statistical accuracy as measurements at the resonance peak.

\bigskip
\noindent{\bf{Measurement of $\Gamma$ using the ratio of the
$\ell^+\ell^-$ final state and total cross sections}} -
First of all let us discuss
 the measurement of absolute and relative branching ratios. This is
possible by simply measuring the cross sections in different
final states, holding the energy fixed:
$\sigma_c^{F_1}/\sigma_c^{F_2}=B_{F_1}/B_{F_2}$
and $\sigma_c^{F}/\sigma_c=B_{F}$.
As apparent from Eq.~(\ref{ratio}), such cross section ratios do not depend on
$\sigma_M$. Therefore, we can measure
branching ratios and ratios of branching ratios
with no systematic error induced by the energy spread. Of course,
to measure $\sigma_c$, we must be able to sum over all final
states for which $B_{F}$ is substantial, taking into account
backgrounds and systematics associated with
correctly determining the relative normalizations
of different final states.  This might not be easy, and could
even be impossible if the resonance has some effectively invisible decays
unless the branching ratio for invisible decays can be measured
in a different experimental setting (in particular, one
in which invisible decays can be effectively `tagged'
by producing the resonance in association with some other particle).

Given $B_{\ell^+\ell^-}$, the value
of $\sigma_c$ and Eq.~(\ref{scaled-conv3})
can then be used to determine $\Gamma$ if $\sigma_M$
is known.  The error on $\Gamma$
induced by uncertainty in $\sigma_M$, given the absence of systematic error in
$B_{\ell^+\ell^-}$, is obtained from Eq. (\ref{condition2}) with $\Delta B=0$,
\begin{equation}\label{condition3}
\Phi(x,\gamma)=\Phi(x+\Delta x, \gamma+\Delta\gamma)\,.
\end{equation}
If we measure $\sigma_c$ and
$\sigma_c^{\ell^+\ell^-}$ at the resonance peak we have $\Delta x=0$ and
Eq.~(\ref{condition3}) then requires $\Delta\gamma=0$; in turn,
\begin{equation}
\Delta\gamma=0\longrightarrow \frac{\Delta\Gamma}{\Gamma}=\frac
{\Delta\sigma_M}{\sigma_M}\,.
\end{equation}
Therefore, the systematic error induced in $\Gamma$
does not depend on $\gamma$. On the other
hand, if we move away from the peak, the situation changes
drastically, as illustrated in Fig. \ref{fig3}  which gives
$\Delta\Gamma/\Gamma$ vs. $\gamma$ for $\Delta\sigma_M/\sigma_M$
equal to 0.05. This figure can be trivially scaled for different
values of $\Delta\sigma_M/\sigma_M$. We see that above
$\gamma\approx 2$ the error decreases as one moves further away from the
resonance peak. For values below $\gamma\approx 2$ the situation is more
complicated, but for each choice of the energy there is a zero in
the error.~\footnote{A really precise computation
of the locations of these zeroes would need to include
the effects of bremsstrahlung.}
Therefore, by opportunely choosing the energy of the
measurement one could try to work in a region where the
systematic error is
very much reduced. For instance, if  $\Gamma\approx 1.4\sigma_M$,
by taking $E=M+\Gamma$, one has $\Delta\Gamma=0$. Notice that the
locations of the zeroes do not depend on $\Delta\sigma_M$.

There are two potentially severe difficulties associated with
this technique.  First, in many instances the $\ell^+\ell^-$
decay mode has a small branching ratio. If the resulting
event rate in the $\ell^+\ell^-$ final state is not large, the statistical
error for $\sigma_c^{\ell^+\ell^-}$ will be large. Statistics for $E$
significantly different from $M$ (as required to take advantage of
the zeroes discussed above) would be the first to become problematical.
Second, we have already noted that
measurement of $\sigma_c$ may be quite tricky. One must
correctly normalize different final state channels relative to one
another and assume that all final state channels
with substantial branching ratio are visible.
Regarding the latter, there is the alternative of measuring
the branching ratio for invisible final states
using other experimental situations/techniques and then making the appropriate
correction to compute the full $\sigma_c$ given
the contributions to $\sum_{F}\sigma_c^{F}$ that
can be measured in the $s$-channel setting.

Of course, in
some instances $B_{\ell^+\ell^-}$ will also be measured at other
machines or via other processes. Also in this case, one can immediately
determine $\Gamma$ given a measurement of $\sigma_c$.  Alternatively,
if $B_{F}$ is also known for some final state $F$ (as well as
$B_{\ell^+\ell^-}$), $\sigma_c^F/(B_{\ell^+\ell^-}B_F)$ will
give a determination of $\Gamma$ with systematic uncertainty
from $\Delta \sigma_M$ as described above.

\bigskip

\noindent {\bf{Measurement of the Breit-Wigner area}} - As apparent from
Eq. (\ref{integral}), the energy integral of the cross section
does not depend on $\sigma_M$, and, in the narrow width
approximation, is given by
\begin{equation}\label{integral2}
\int\sigma_c(E)\,dE= 4\pi^2(2j+1)\,\Gamma\cdot\,B_{\ell^+\ell^-}
\end{equation}
Therefore, even if the energy spread is very poorly known, 
so long as it does not change as one scans over the resonance one
could still measure the integral of the cross section, which is
proportional to the product $\Gamma\cdot B_{\ell^+\ell^-}$,
and obtain $B_{\ell^+\ell^-}$ from the ratio of the peak cross sections
$\sigma_c^{\ell^+\ell^-}/\sigma_c$
(or, possibly, in another experimental setting). Of course, as discussed
with regard to the previous procedure, the statistical error for
$\sigma_c^{\ell^+\ell^-}$ might be large
even when measured at the resonance peak. In addition,
the event rate substantially off-resonance (as compared
to both $\Gamma$ and $\sigma_M$) would be small
(while typical backgrounds would be essentially constant).
Thus, one can often only measure the cross section with
good statistical accuracy over a limited
portion of the full range.

If $\Gamma\gg\sigma_M$, and $\sigma_M$ is known ahead of time
(via spin-precession measurements or the like), this fact will
quickly become apparent after measuring the cross section at a few
energy settings with $E-M\gg\sigma_M$.
A reliable value for $\Gamma\cdot\,B_{\ell^+\ell^-}$
can normally be obtained so long as statistics are good at the peak.
The procedure is that based on Eq.~(\ref{largegam}).
One will measure $\sigma_c(E)$
for a set of points perhaps out to $E-M\sim \pm 2\Gamma$,
and the remainder of the integral will be determined using
the Breit-Wigner shape that would have been revealed by
the measurements made. Sensitivity to $\sigma_M$ will be
very minimal.

If $\sigma_M\gg\Gamma$, then the energy dependence of $\sigma_c(E)$
is determined by $\sigma_M$, see Eq.~(\ref{smallgam}),
or more generally by $g$ of Eq.~(\ref{scaling})
(bremsstrahlung should be included). Given a known value for $\sigma_M$
and a set of measured points, the remainder of the integral can be computed.
However, statistical errors are likely to be quite large because
the cross section is suppressed by the ratio $\Gamma/\sigma_M$.

Clearly, the most difficult case is that 
in which the resonance is very narrow and
the smallest achievable $\sigma_M$ value is such
that $\Gamma\sim\sigma_M$.
Unless statistics remain good far off the peak,
which is not likely,
deconvolution of the effects of $\sigma_M$ and $\Gamma$ is required.
The simplest deconvolution procedures for known $\sigma_M$ are the
scan and ratio procedures outlined previously.
Thus, if $\Gamma\sim \sigma_M$, the three-point scan technique
and the technique of varying $\sigma_M$ while keeping $E=M$ are
the most efficient for determining the parameters of an
$s$-channel resonance so long as systematic uncertainty in the energy spread
of the beam is smaller than $1\%\div 2\%$.
(Of course, the appropriate strategy for exploring a resonance
would be quite different if $\sigma_M$ is not independently
measured with high statistical accuracy using the precession measurements.)


\bigskip
\section{Applications}
\begin{figure}[p]
\epsfysize=16truecm
\centerline{\epsffile{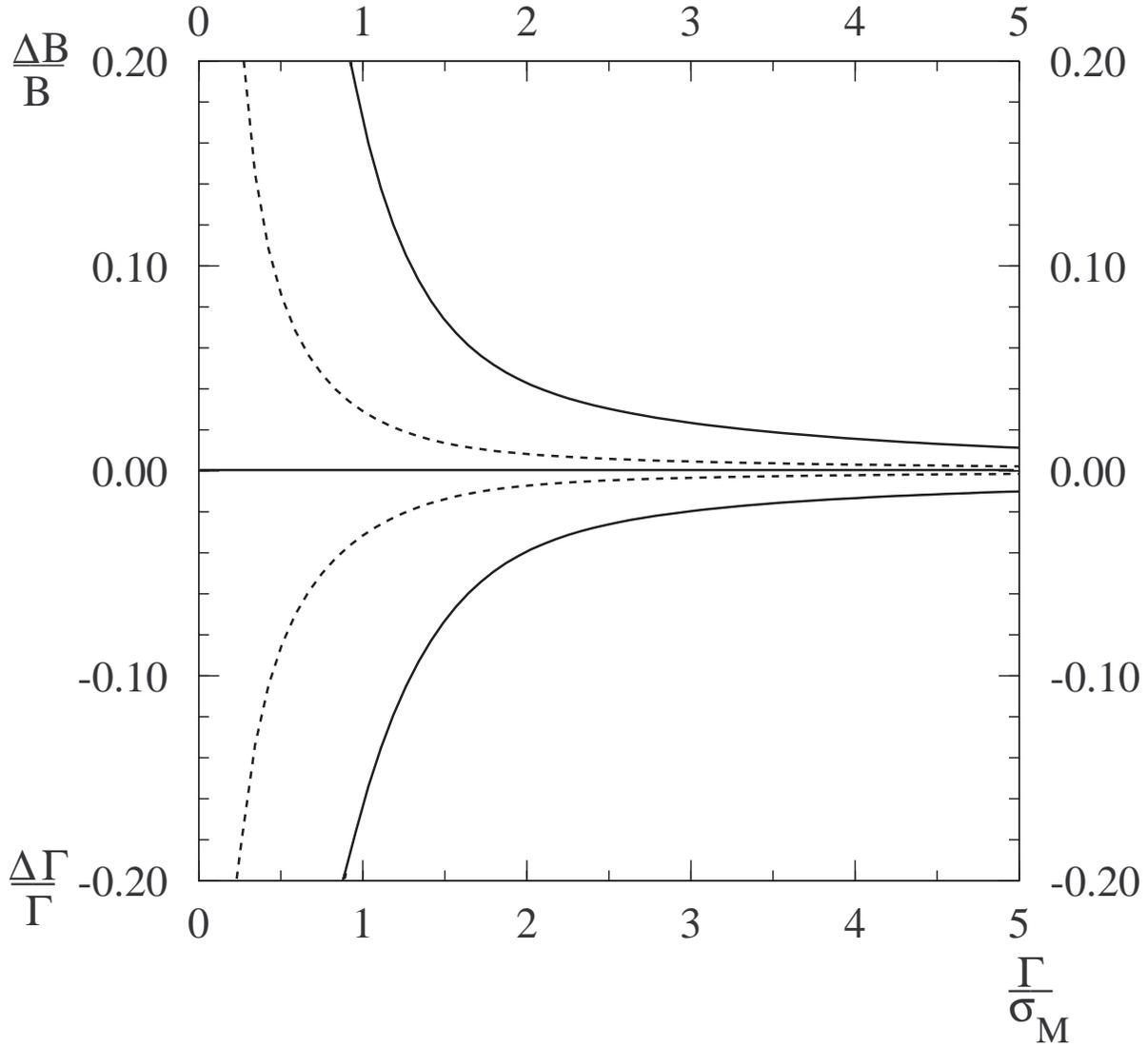}}
\noindent
\caption{
The fractional errors induced in $B$ and $\Gamma$
by $\Delta\sigma_M$ as functions of
$\Gamma/\sigma_M$ when employing the three-point scan  measurement ($E=M$,
$E=M\pm 2\Gamma$); results are shown for $\Delta\sigma_M/\sigma_M=0.05$ (solid
line) and $\Delta\sigma_M/\sigma_M=0.01$ (dashed line).}
\label{fig2}
\end{figure}

\begin{figure}[p]
\epsfysize=14truecm
\centerline{\epsffile{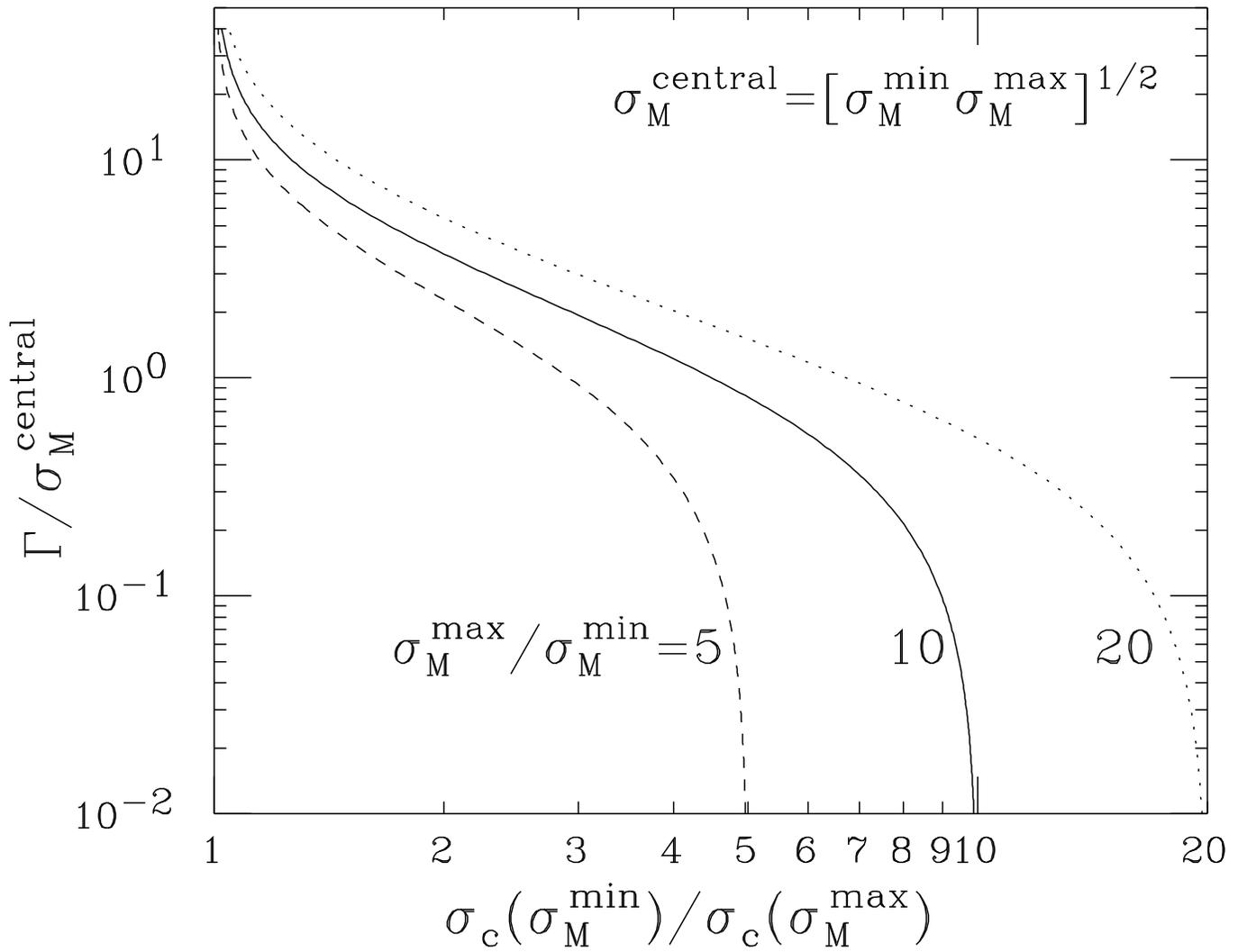}}
\noindent
\caption{
$\Gamma/\sigma_M^{\rm central}$ plotted as a function of the cross section
ratio $\sigma_c(\sigma_M^{\rm min})/\sigma_c(\sigma_M^{\rm max})$
for various values of $\sigma_M^{\rm max}/\sigma_M^{\rm min}$ keeping
$\sigma_M^{\rm central}\equiv\protect
\sqrt{\sigma_M^{\rm max}\sigma_M^{\rm min}}$ fixed.}
\label{dlgdlr}
\end{figure}

\begin{figure}[p]
\epsfysize=16truecm
\centerline{\epsffile{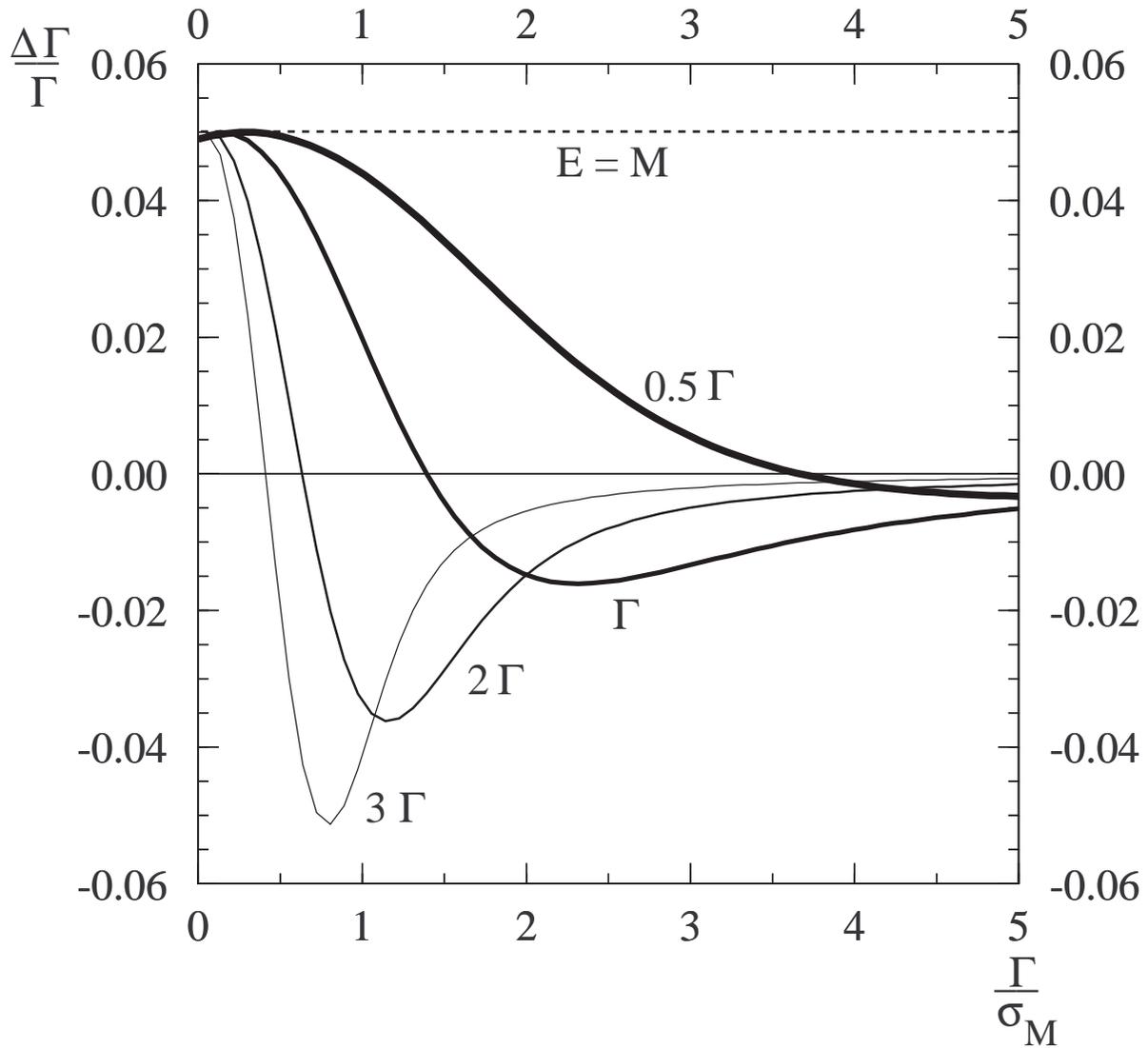}}
\noindent
\caption{
The fractional error induced in  $\Gamma$  as a function of
$\Gamma/\sigma_M$ for $E-M=k\Gamma$, with $k=0,1,2,3$, assuming
a known value of $B_{\ell^+\ell^-}$ from the
peak measurements of $\sigma_c$ and $\sigma_c^{\ell^+\ell^-}$.  The
value of $\Delta\sigma_M/\sigma_M$ is fixed at 0.05.}
\label{fig3}
\end{figure}

\begin{figure}[p]
\epsfysize=16truecm
\centerline{\epsffile{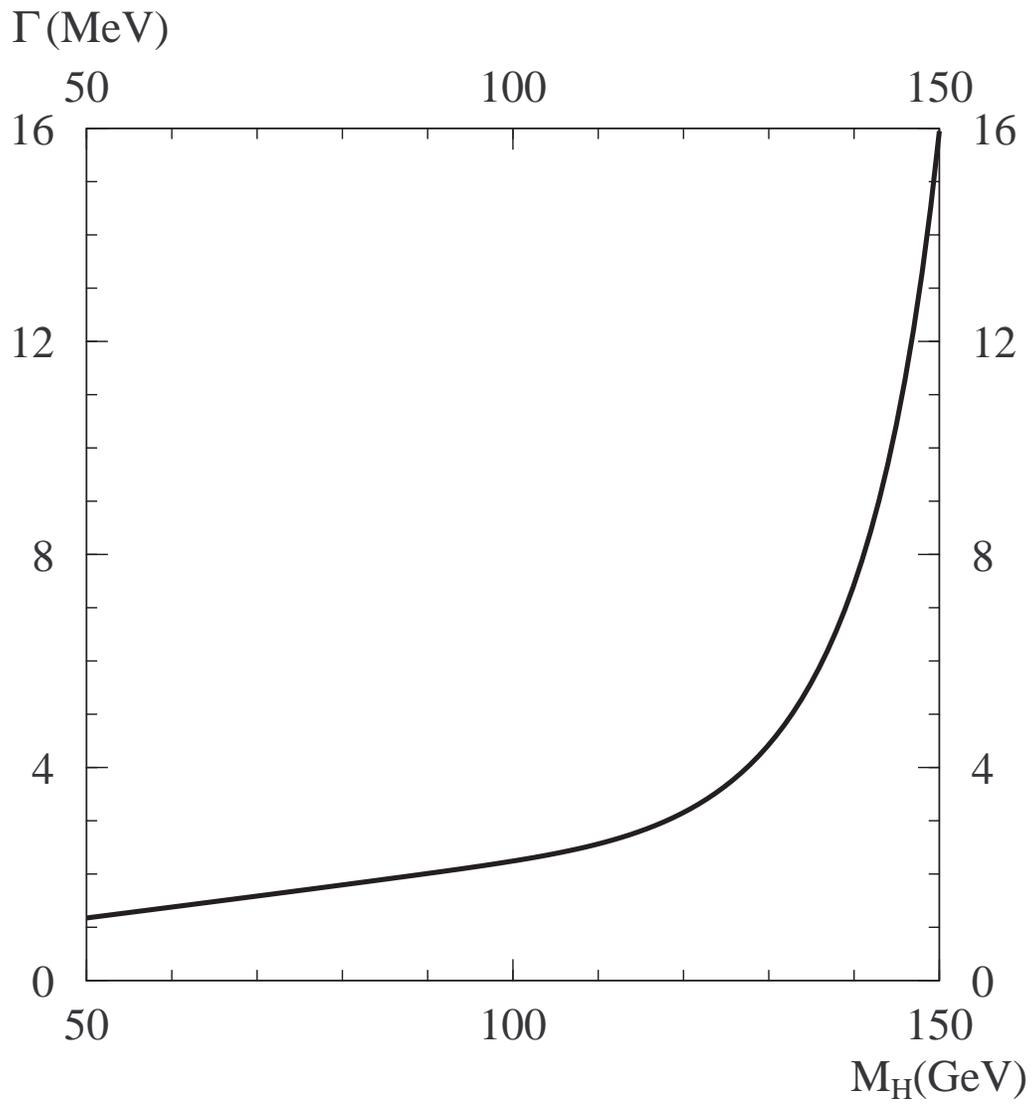}}
\noindent
\caption{
The total width of the SM-Higgs boson as a function of $M_H$.}
\label{fig4}
\end{figure}

\begin{figure}[p]
\epsfysize=16truecm
\centerline{\epsffile{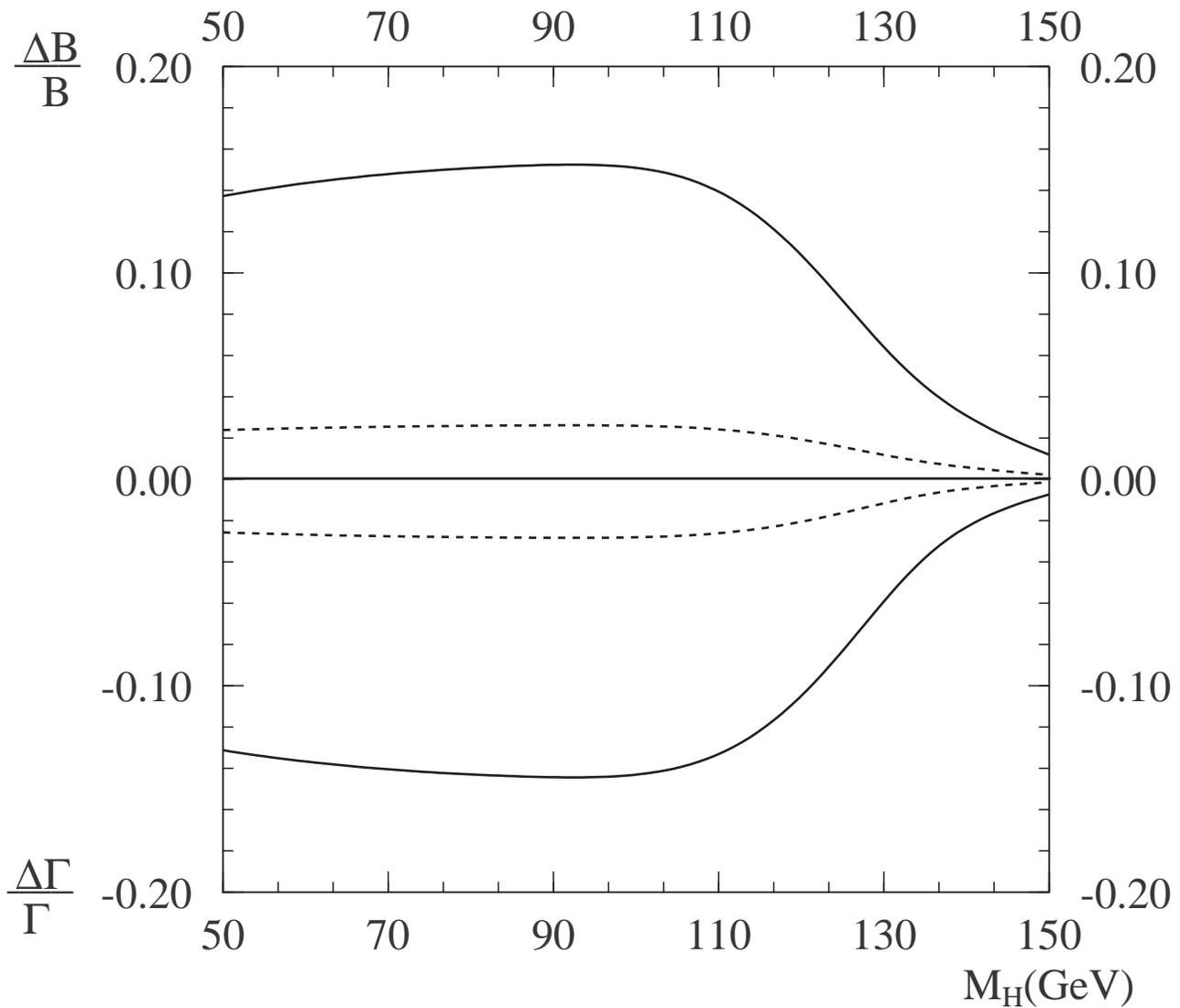}}
\noindent
\caption{
The $\Delta\sigma_M$-induced fractional errors in $B$ and $\Gamma$
as functions of $M_H$ assuming a three-point resonance scan ($E=M$, $E=M\pm
2\Gamma$); results are shown for
$\Delta\sigma_M/\sigma_M=0.05$ (solid line) and
$\Delta\sigma_M/\sigma_M=0.01$ (dashed line). The value of the
beam resolution $R$ has been fixed to 0.003\%.}
\label{fig5}
\end{figure}

\bigskip
\noindent {\bf{SM-Higgs boson}} - In
this Section we will apply the first two methods of Section
\ref{general} to the study of a light SM-Higgs boson. 
Consider first the three-point scan. By using
the results of Fig. \ref{fig2} one can easily determine the errors induced
by the uncertainty in the energy spread as a function of the Higgs
mass. For the evaluation of the Higgs width as a function of the
mass, we make use of the expressions given in Ref.
\cite{djouadi}. The decay channels we have considered are $b\bar
b$, $\tau^+\tau^-$, $c\bar c$, $gg$ and $WW^*$, $ZZ^*$ (one
of the two final bosons being virtual). The resulting total width
is given in Fig. \ref{fig4}. We have considered the interval $50\le
M_H({\rm GeV})\le 150$, but recall that, actually, there is an
experimental lower bound at $90\%$ C.L. of about 90 GeV
\cite{vancouver}. Below 110 GeV, the width of the  Higgs
increases approximately linearly with the mass (aside from logarithmic
effects due to the running of the quark masses) which means that the
ratio $\Gamma_H/\sigma_M$ is approximately constant. By choosing
$R=0.003\%$ (see \cite{gunmumu}) we get $\Gamma_H/\sigma_M\approx
1$. From Fig. \ref{fig2}, we see that the $\Delta\sigma_M$-induced errors
in $B$ and in $\Gamma_H$ are
about 15\% and 2.5\% for $\Delta\sigma_M/\sigma_M=0.05$ and 0.01,
respectively, for a $k=2$ scan (as appropriate, given that
$\Gamma_H\sim\sigma_M$, for comparing
to the $E=M,~E=M\pm 2\sigma_M$ scan summarized below 
that was used to estimate statistical errors).
Above 110 GeV, the systematic errors decrease rapidly due to
the fast increase of the width. Fig. \ref{fig5} summarizes
the $\Delta B/B$ and $\Delta \Gamma_H/\Gamma_H$ fractional 
systematic errors as a function of $M_H$ for $R=0.003\%$ and $k=2$.
Notice that, at least in the region up to 110 GeV, it will be
mandatory to have values of $R$  of the order 0.003\%. For instance,
if we take $R=0.01\%$, corresponding to
$\Gamma_H/\sigma_M\approx 0.3$ (for $50\le M_H({\rm GeV})\le 110$), the
fractional $\Delta\sigma_M$-induced errors in $B$ and $\Gamma_H$
increase to about $14\%\div 16\%$ for $\Delta\sigma_M/\sigma_M=0.01$,
as can be seen from the dashed line of Fig. \ref{fig2}.

\begin{table}[h]
\caption[fake]{\baselineskip 0pt Percentage errors ($1\sigma$) for 
$\sigma_cB(H\to b\anti b,W\wstar,Z\zstar)$
(extracted from channel rates)
and $\Gamma_H$ for $s$-channel Higgs production at the muon collider
assuming beam energy resolution of $R=0.003\%$. Results are
presented for two integrated four-year luminosities:
$L=4\fbi$ ($L=0.4\fbi$). An optimized three-point scan
is employed using measurements at $E=M_H$,
$E=M_H+2\sigma_M$ and $E=M_H-2\sigma_M$, with luminosities of $L/5$,
$2L/5$ and $2L/5$, respectively. [For the cross section measurements, 
this is equivalent to $L\sim 2\fbi$  ($L=0.2\fbi$) at the $E=M_H$ peak].
This table is taken from Ref.~\cite{gunmumu}. Efficiencies and cuts
are those employed in \cite{bbgh}. The effects of bremsstrahlung are
included.}
\small
\begin{center}
\begin{tabular}{|c|c|c|c|c|}
\hline
 Quantity & \multicolumn{4}{c|}{Errors} \\
\hline
\hline
 {\bf Mass (GeV)} & {\bf 80} & {\bf $\mz$} & {\bf 100} & {\bf 110} \\
\hline
$\sigma_cB(b\anti b)$ & $0.8\%(2.4\%)$ & $7\%(21\%)$ & $ 1.3\%(4\%)$ & 
$ 1\%(3\%)$ \\
\hline
$\sigma_cB(W\wstar)$ & $-$ & $-$ & $ 10\%(32\%)$ & 
$ 5\%(15\%)$ \\
\hline
$\sigma_cB(Z\zstar)$ & $-$ & $-$ & $-$ & 
$ 62\%(190\%)$ \\
\hline
 $\Gamma_H$ & $ 3\%(10\%)$ & $ 25\%(78\%)$ & $ 10\%(30\%)$ & 
  $ 5\%(16\%)$ \\
\hline
\hline
 {\bf Mass (GeV)} & {\bf 120} & {\bf 130} & {\bf 140} & {\bf 150} \\
\hline
$\sigma_cB(b\anti b)$ & $1\%(3\%)$ & $1.5\%(5\%)$ & $ 3\%(9\%)$ & 
$ 9\%(28\%)$ \\
\hline
$\sigma_cB(W\wstar)$ & $3\%(10\%)$ & $2.5\%(8\%)$ & $ 2.3\%(7\%)$ & 
$ 3\%(9\%)$ \\
\hline
$\sigma_cB(Z\zstar)$ & $16\%(50\%)$ & $10\%(30\%)$ & $ 8\%(26\%)$ & 
$ 11\%(34\%)$ \\
\hline
 $\Gamma_H$ & $ 5\%(16\%)$ & $ 6\%(18\%)$ & $ 9\%(29\%)$ &
  $ 34\%(105\%)$ \\
\hline
\end{tabular}
\end{center}
\label{fmcerrors}
\end{table}

It is interesting to compare these systematic errors to the statistical
errors. The analysis at a muon collider done in Ref. \cite{gunmumu}
gives statistical errors for a three-point scan using scan points
at $E=M,~E=M\pm 2\sigma_M$ and $R=0.003\%$, assuming
$L=4\fbi$ or $L=0.4\fbi$ total accumulated luminosity
(corresponding to 4 years of operation for optimistic or pessimistic,
respectively, instantaneous luminosity). The results
of that analysis are summarized in Table~\ref{fmcerrors}.
Except for $M_H\sim M_Z$, the $L=4\fbi$ 
statistical error for measuring the total width would be of order 
$3\%\div 10\%$ when $80\le M_H\le 140\gev$.  Therefore, to avoid contaminating
this high precision measurement error with systematic
uncertainty from $\Delta\sigma_M$ we will certainly want
to have $\Delta\sigma_M/\sigma_M\lsim 1\%$.
If one adopts the $L=0.4\fbi$
luminosity assumption (the benchmark value of 
Refs.~\cite{reportusa,reporteurope}),
and if $m_H<130\gev$ and not near $M_Z$, 
the statistical measurement error
for $\Gamma_H$ is in the 10\% to 20\% range. This means 
that $\Delta\sigma_M/\sigma_M$ as large as $5\%$ would
be very undesirable. 
Finally, we re-emphasize the fact that performing the
scan using larger $R$ than $R=0.003\%$ leads to larger
statistical errors until $\mh$ approaches the $WW$ decay
threshold and $\gamh/\sigm(R=0.003\%)>5$. For lower $\mh$,
the $R=0.003\%$ results are the best
that can be achieved despite the smaller luminosity at $R=0.003\%$
as compared to higher $R$ values. For example, the error in $\gamh$
for a given luminosity using $R=0.01\%$ 
can be read off from Fig.~13 of \cite{bbgh}. One finds that
$L(R=0.01\%)/L(R=0.003\%)=20,10,2$ is required in order that
the $\gamh$ statistical errors for $R=0.01\%$ be equal to those
for $R=0.003\%$ at $\mh=130,140,150\gev$, respectively. Existing
machine designs are such that 
$\lyear(R=0.01\%)/\lyear(R=0.003\%)=0.22\fbi/0.1\fbi=2.2$.  Thus,
increasing $R$ would not improve the scan-procedure statistical errors 
until $\mh>150\gev$. In addition, $\Delta\sigm$-induced
systematic errors always rise rapidly with increasing $R$. For
$\mh\leq 150\gev$, one should employ the smallest value of $R$ possible.

\begin{table}[h]
\caption[fake]{\baselineskip 0pt Percentage errors ($1\sigma$) for 
$\sigma_cB(H\to b\anti b,W\wstar,Z\zstar)$
(extracted from channel rates)
and $\Gamma_H$ for $s$-channel Higgs production at the muon collider.
Operation at $E=M_H$ is assumed and
the $r_c$-ratio technique is used for determining $\Gamma_H$.
Results are presented assuming accumulated luminosities of
$L=2\fbi$ ($L=0.2\fbi$) at $R=0.003\%$
and $L=9.4\fbi$ ($L=0.94\fbi$) at $R=0.03\%$, corresponding
to roughly two years of running at each $R$ for optimistic (pessimistic) 
instantaneous luminosity assumptions. Efficiencies and cuts
employed appear in \cite{bbgh}. The effects of bremsstrahlung are included.}
\small
\begin{center}
\begin{tabular}{|c|c|c|c|c|}
\hline
 Quantity & \multicolumn{4}{c|}{Errors} \\
\hline
\hline
 {\bf Mass (GeV)} & {\bf 80} & {\bf $\mz$} & {\bf 100} & {\bf 110} \\
\hline
$\sigma_cB(b\anti b)$ & $0.7\%(2.2\%)$ & $6.3\%(20\%)$ & $ 1.2\%(3.8\%)$ & 
$ 0.9\%(2.8\%)$ \\
\hline
$\sigma_cB(W\wstar)$ & $-$ & $-$ & $ 8.2\%(26\%)$ & 
$ 3.8\%(12\%)$ \\
\hline
$\sigma_cB(Z\zstar)$ & $-$ & $-$ & $-$ & 
$ 60\%(190\%)$ \\
\hline
 $\Gamma_H$ & $ 6\%(19\%)$ & $ 63\%(200\%)$ & $ 14\%(45\%)$ & 
  $ 8\%(25\%)$ \\
\hline
\hline
 {\bf Mass (GeV)} & {\bf 120} & {\bf 130} & {\bf 140} & {\bf 150} \\
\hline
$\sigma_cB(b\anti b)$ & $0.9\%(2.8\%)$ & $1.4\%(4.4\%)$ & $ 2.4\%(7.6\%)$ & 
$ 6.6\%(21\%)$ \\
\hline
$\sigma_cB(W\wstar)$ & $2.4\%(7.7\%)$ & $1.8\%(5.7\%)$ & $ 1.6\%(5.0\%)$ & 
$ 1.8\%(5.6\%)$ \\
\hline
$\sigma_cB(Z\zstar)$ & $15\%(46\%)$ & $7.9\%(25\%)$ & $ 6.3\%(20\%)$ & 
$ 7.0\%(22\%)$ \\
\hline
 $\Gamma_H$ & $ 6.3\%(20\%)$ & $ 6\%(19\%)$ & $ 5.4\%(17\%)$ &
  $ 4.7\%(18\%)$ \\
\hline
\end{tabular}
\end{center}
\label{fmcerrorsrc}
\end{table}

Let us now compare to the $r_c$-ratio technique. 
We employ the same total of 4 years of operation
as considered for the three-point scan, but always with $E=M_H$.
As noted earlier, a good sharing of time is to devote two years
to running at $R=0.003\%$, accumulating $L=2\fbi$  ($L=0.2\fbi$)
for optimistic (pessimistic) instantaneous luminosity, and a
second two years to running at $R=0.03\%$, corresponding to [using
the luminosity scaling law of Eq.~(\ref{lform})] $L=9.4\fbi$ ($L=0.94\fbi$)
of accumulated luminosity. For $R=0.03\%$, 
$\Gamma_H/\sigma_M^{\rm max}\sim 0.1$ and, as noted earlier, each 
$\sigma_c B_F$ cross section is decreased by almost
the same factor by which the luminosity has increased, leaving
the number of signal events unchanged.
However, the background is increased by a factor of 4.7.
A complete calculation is required.  This was performed in \cite{ratio}.
The various $\sigma_c B_F$ statistical errors 
are summarized in Table~\ref{fmcerrorsrc} along with the
$\Delta\gamh/\gamh$ statistical error computed by combining all the
listed final state channels and following the procedure of
Eq.~(\ref{sigcerror}).
We observe that the ratio technique becomes superior to the scan
technique for the larger $\mh$ values ($\mh>130\gev$). This
is correlated with the fact that $\gamh/\sigmmin$ (where $\sigmmin$
is that for $R=0.003\%$) 
becomes substantially larger than 1 for such $\mh$. In
particular, for larger $\mh$,
$\gamh/\sigm^{\rm central}$ is in a range such that $|s|$
and, consequently, the error in $\gamh$ will be minimal.
Thus, the two techniques are actually quite complementary --- by employing
the best of the two procedures, a very reasonable determination of $\gamh$
and very precise determinations of the larger channel rates
will be possible for all $\mh$ below $2\mw$.

Finally, we again note that the $r_c$-ratio technique
has the advantage that the $\Delta\sigma_M$-induced systematic error  
in $\Gamma_H$ is equal to $\Delta\sigma_M/\sigma_M$ and will, therefore,
be smaller by a factor of about 2.5 (if $M_H\leq 120\gev$)
for the $r_c$-ratio technique than 
for the scan technique and that there is no $\Delta\sigma_M$-induced
systematic error in the $B_{\ell^+\ell^-}B_F$ determinations.
Thus, although the statistical errors for the $r_c$-ratio
procedure are larger than for the scan procedure when $\mh<130\gev$,
if the fractional error in $\sigma_M$ is substantially larger
than $0.01$, the $r_c$-ratio could have net overall (statistical
plus systematic) error that is smaller than the scan technique
down to $\mh$ values significantly below $130\gev$.

\bigskip

\noindent {\bf{The lightest PNGB}} - The $s$-channel
production of the lightest neutral
pseudo-Nambu-Goldstone boson (PNGB) ($P^0$), present in models of
dynamical breaking of the electroweak symmetry which have a
chiral symmetry larger than $SU(2)\times SU(2)$, has recently been
explored \cite{mumu,mumualso}. In the broad class
of models considered in \cite{mumu}, the $\pzero$ is of particular interest
because it contains only down-type techniquarks (and charged technileptons)
and thus has a mass scale that is most naturally  set by the
mass of the $b$-quark. Other color-singlet PNGB's will have masses
most naturally set by $\mt$, while color non-singlet PNGB's will
generally be even heavier. The $\mpzero$ mass range, that is typically
suggested by technicolor models \cite{technicolor}, is
$10\gev<\mpzero<200\gev$.

Discovery of the $\pzero$ in the $gg\to\pzero\to\gam\gam$
mode at the Tevatron Run II and at the
LHC will almost certainly be possible unless its
mass is either very small ($\lsim 30\gev$?) or very large ($\gsim
200\gev$?), where the question marks are related to uncertainties in
backgrounds in the inclusive $\gam\gam$ channel.
Run I data at Tevatron can already be used to exclude a $\pzero$ in the
$50-200\gev$ mass range for a number of technicolors $\ntc> 12-16$.
In contrast, an $\epem$ collider, while able to discover the $\pzero$
via $\epem\to\gam\pzero$, so long as $\mpzero$ is not close to $\mz$,
is unlikely (unless the TESLA $500\fbi$ per year option
is built or $\ntc$ is very large) to be able to determine the rates for
individual $\gam F$ final states ($F=b\anti b,\tauptaum,gg$
being the dominant $\pzero$ decay modes) with sufficient accuracy such as
to yield more than very rough indications about the
parameters of the technicolor model. The $\gam\gam$ collider mode of
operation at an $\epem$ collider would allow one to discover and study
the $\pzero$ with greater precision.

\begin{figure}[p]
\epsfysize=16truecm
\centerline{\epsffile{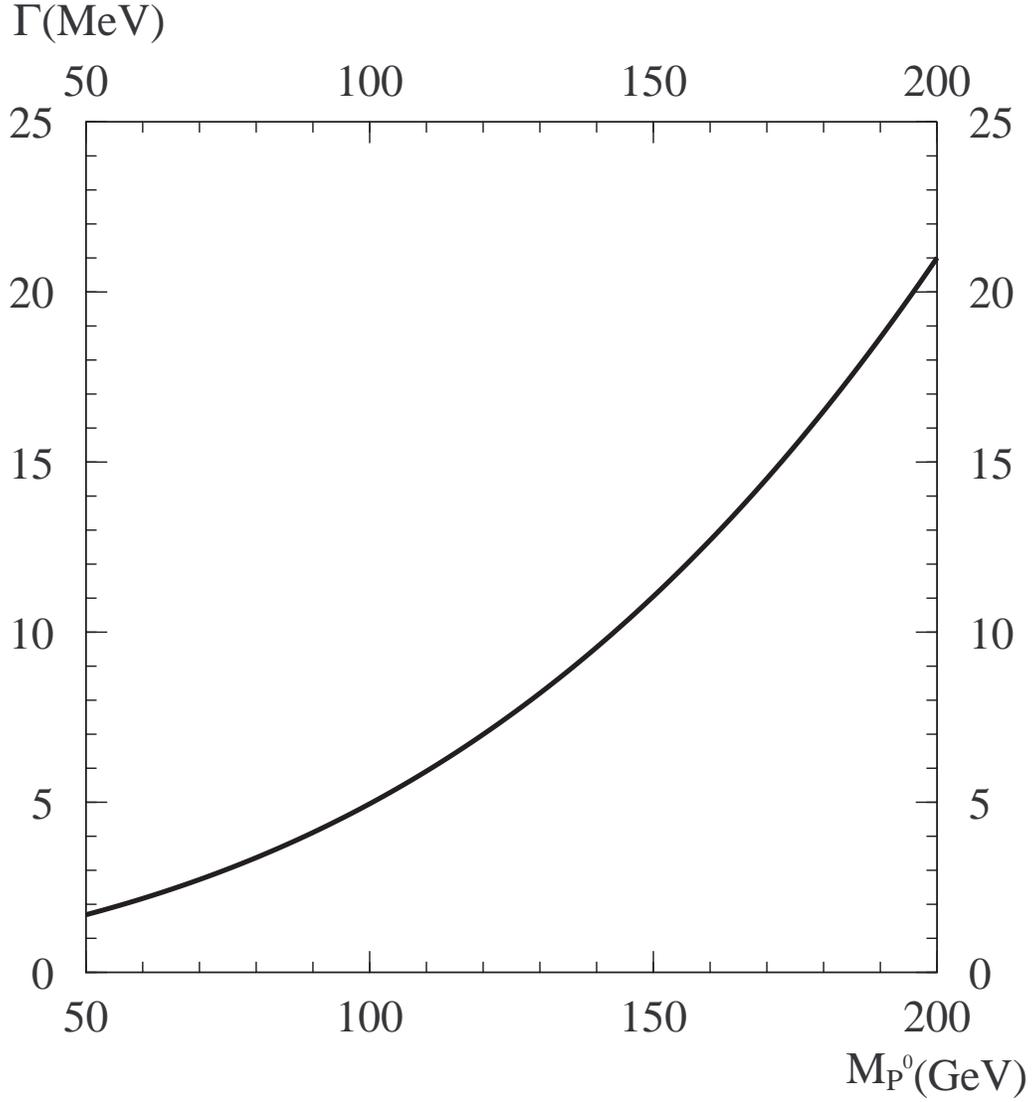}}
\noindent
\caption{
The total width of the lightest PNGB $P^0$ as a function of $M_{P^0}$
(for the choice of model parameters given in Ref. \protect\cite{mumu}).}
\label{fig6}
\end{figure}

A $\mupmum$ collider would play a very special role
with regard to determining key properties of the $\pzero$.
In particular, the $\pzero$, being comprised
of $D\anti D$ and $E\anti E$ techniquarks, will naturally have couplings
to the down-type quarks and charged leptons of the SM.
Thus, $s$-channel production ($\mupmum\to\pzero$) is predicted
to have a substantial rate for $\rts\sim\mpzero$.
Because the $\pzero$ has a very narrow width (see
Fig. \ref{fig6}), not much
larger than that of a SM-Higgs boson of the same mass,
in order to maximize this rate
it is important that one operates the $\mupmum$ collider
so as to have extremely small beam energy spread, $R=0.003\%$.
For such an $R$, the resolution in $E=\rts$ of the muon collider,
$\sigma_E$, is of order $\sigma_E\sim 1\mev (E/50\gev)$, whereas
the $\pzero$ width, $\gampzero$, varies from $2\mev$ to $20\mev$ as $\mpzero$
ranges from $50\gev$ up to $200\gev$ (small differences with
respect to \cite{mumu} come from running fermion masses). Thus,
$\sigma_E<\gampzero$ is possible and leads to very high $\pzero$
production rates for typical $\mupmum\to\pzero$ coupling strength.

Assuming that the $\pzero$ is discovered at the Tevatron, the LHC
or (as might be the only possibility if $\mpzero$ is very small)
at an $\epem$ collider (possibly operating in the $\gam\gam$
collider mode), the $\mupmum$ collider could quickly (in less
than a year) scan the mass range indicated by the previous
discovery (for the expected uncertainty in the mass
determination) and center on $\rts\simeq\mpzero$ to within
$<\sigma_M$. A first very rough estimate of $\gampzero$
would also emerge from this initial scan. One would then proceed
with a dedicated study of the $\pzero$. One technique would be
to use the optimal three-point scan \cite{mumu}
of the $\pzero$ resonance (with measurements
at $E=\mpzero$ and $E=\mpzero\pm 2\sigma_M$
using $R=0.003\%$). The three-point scan would determine with high
statistical precision all the $\mupmum\to\pzero\to F$ channel
rates and give a reasonably accurate measurement of
the total width $\gampzero$. For the particular
technicolor model parameters analysed in  \cite{mumu}, 4 years
of the pessimistic yearly  luminosity ($L_{\rm year}=0.1\fbi$)
devoted to the scan yields the results presented
in Fig.~19 of \cite{mumu}.~\footnote{This figure gives the errors
before taking into account the possible variation of
the luminosity with $\mpzero$. If the muon collider
is built so that the $\sqrt s=100\gev$ luminosity is maximized,
then $L_{\rm year}$ will be smaller (larger) 
than the $\sqrt s=100\gev$ value for smaller (larger) $\sqrt s$.
The relevant luminosity scaling laws are those given in
Eq.~(7.2) of Ref.~\cite{mumu}. The effects upon the statistical errors
quoted below of such luminosity scaling are given in
Fig.~20 of \cite{mumu}, but will not be included in our discussion here.}
Sample statistical errors for 
$\sigma_cB(\pzero\to {\rm all})$ and $\gampzero$ taken from this figure
at $\mpzero=60\gev$, $80\gev$, $\mz$, $110\gev$, $150\gev$ and $200\gev$
are given in Table~\ref{fmcerrorspgb}.

\begin{table}[h]
\caption[fake]{\baselineskip 0pt Fractional 
statistical errors ($1\sigma$) for 
$\sigma_cB(\pzero\to {\rm all})$ (combining $b\bar b$, $\tauptaum$,
$c\bar c$ and $gg$ tagged-channel rates --- see \cite{mumu})
and $\gampzero$ for $s$-channel $\pzero$ production at the muon collider.
We compare results for an $R=0.003\%$ three-point scan
with total integrated luminosity of $L=0.4\fbi$ (corresponding
to four years of running at pessimistic luminosity, with 
distribution $L/5$ at $E=\mpzero$,
$2L/5$ at $E=\mpzero+2\sigma_M$ and $2L/5$ at $E=\mpzero-2\sigma_M$)
to results obtained using the $r_c$-ratio technique
assuming accumulated luminosities at $E=\mpzero$
of $L=0.2f\fbi$ at $R=0.003\%$ and 
$L=0.94(2-f)\fbi$ at $R=0.03\%$ (corresponding
to roughly $2f$ years of running at $R=0.003\%$ and $(4-2f)$
years of running at $R=0.03\%$ for pessimistic
instantaneous luminosity assumptions). $f$ (tabulated below) is chosen
to minimize the error in $\gampzero$. We employ the 
efficiencies, cuts and tagging procedures described in \cite{mumu}.
The effects of bremsstrahlung are included.}
\small
\begin{center}
\begin{tabular}{|c|c|c|c|c|c|c|}
\hline
 Quantity & \multicolumn{6}{c|}{Errors for the scan procedure} \\
\hline
\hline
 {\bf Mass (GeV)} & {\bf 60} & {\bf 80} & {\bf $\mz$} & {\bf 110} 
& {\bf 150} & {\bf 200} \\
\hline
$\sigma_cB$ &  0.0029 & 0.0054 & 0.043 & 0.0093 & 0.012 & 0.018\\
\hline
 $\gampzero$ & 0.014 & 0.029 & 0.25 & 0.042 & 0.052 & 0.10 \\
\hline
\hline
 Quantity & \multicolumn{6}{c|}{Errors for the $r_c$-ratio procedure} \\
\hline
\hline
 {\bf Mass (GeV)} & {\bf 60} & {\bf 80} & {\bf $\mz$} & {\bf 110} 
& {\bf 150} & {\bf 200} \\
\hline
$f$ & 0.8 & 0.7 & 0.6 & 0.8 & 0.9 & 1.0 \\
\hline
$\sigma_cB$ & 0.0029 & 0.0062 & 0.055 & 0.010 & 0.011 & 0.016 \\
\hline
 $\gampzero$ & 0.014 & 0.028 & 0.24 & 0.041 & 0.039 & 0.053 \\
\hline
\end{tabular}
\end{center}
\label{fmcerrorspgb}
\end{table}

\begin{figure}[p]
\epsfysize=16truecm
\centerline{\epsffile{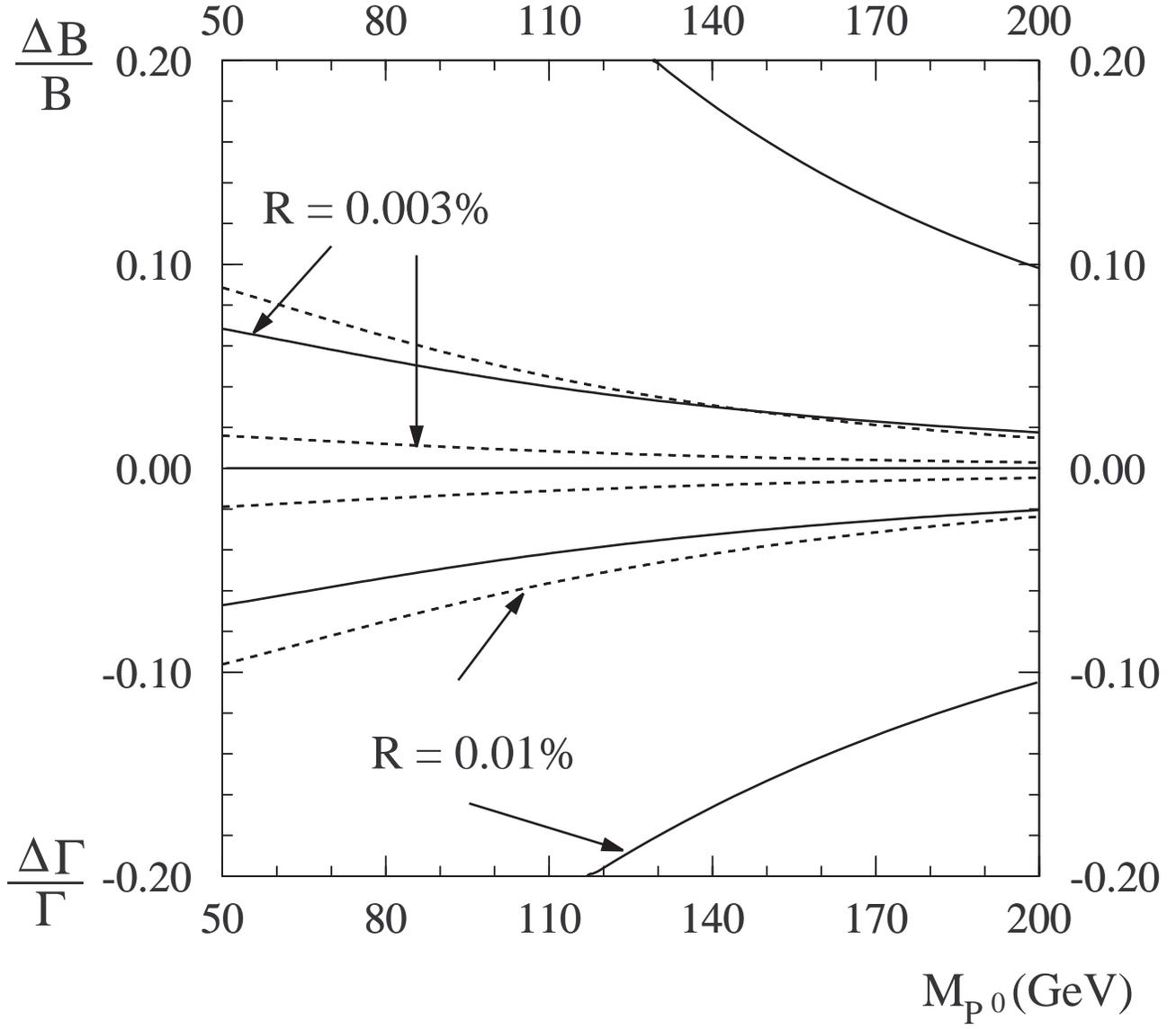}}
\noindent
\caption{
The $\Delta\sigma_M$-induced fractional errors in $B$ and $\Gamma=\gampzero$
as functions of $M_{P^0}$ assuming a three-point resonance scan ($E=M$, $E=M\pm
2\sigma_M$); results are shown for
$\Delta\sigma_M/\sigma_M=0.05$ (solid line) and
$\Delta\sigma_M/\sigma_M=0.01$ (dashed line), 
and for  $R=0.003\%$ and $R=0.01\%$.}
\label{fig7}
\end{figure}

In the analysis performed in \cite{mumu}, 
we did not consider  the errors induced
by a systematic error in the energy spread. In Fig. \ref{fig7} we show
the $\Delta\sigma_M$-induced fractional errors for the branching ratio $B$
(where $B=B_{\ell^+\ell^-}B_F$ if we focus on a given final state
or $B=B_{\ell^+\ell^-}$ if we sum over all final states)
and for $\gampzero$ predicted for the same three-point scan measurement 
($E=M,~E=M\pm 2\sigma_M$, $R=0.003\%$) as employed
for the statistical error analysis summarized above.
Results are shown for the two cases of 
$\Delta \sigma_M/\sigma_M=0.05$ (solid line) and
$\Delta \sigma_M/\sigma_M=0.01$ (dashed line). 
For $R=0.003\%$, we find $\Delta\gampzero/\gampzero\sim
c_{\gampzero} \Delta\sigma_M/\sigma_M$ with $c_{\gampzero}\sim 1.5$ 
for $\mpzero\sim 60\gev$ falling to $c_{\gampzero}\sim 0.45$
for $\mpzero\sim 200\gev$. For $\mpzero\lsim 80\gev$, the result
is that the induced $\Delta\gampzero/\gampzero$
systematic errors are comparable to
the expected statistical errors even for $\Delta \sigma_M/\sigma_M=0.01$.
For example, at $\mpzero=60\gev$ both the systematic error
and the statistical error are of order 1.5\%.
In the neighborhood of the $Z$ peak the errors
from the optimal three-point scan \cite{mumu} are largely
dominated by the $Z$ background and the $\Delta
\sigma_M/\sigma_M$ effect can be neglected if $\Delta\sigma_M/\sigma_M$
is not large. For $\mpzero\sim 150\gev$ ($\sim 200\gev$), 
$c_{\gampzero}\sim 0.6$ ($0.45$), \ie\
$\Delta\gampzero/\gampzero\sim 0.6\Delta\sigma_M/\sigma_M$
($0.45\Delta\sigma_M/\sigma_M$), 
while the statistical $\Delta\gampzero/\gampzero$
error from the three-point scan \cite{mumu} would be of the
order $5\%$ ($10\%$). Thus, for $\Delta \sigma_M/\sigma_M\leq 0.01$ 
the systematic error would be much smaller than the statistical error.
As a result, for $\mpzero\gsim 100\gev$, as far as the $\Delta\sigma_M$-induced
errors for $\gampzero$ are concerned, 
one could consider employing a value of $R$ larger than $0.003\%$.
As an example, $R=0.01\%$ could be chosen. Since $\gampzero/\sigma_M$
is still $\gsim 1$ for this $R$ and $\mpzero\gsim 100\gev$,
we can expect that systematic errors will still be under control.
The actual systematic errors
resulting from an $E=M$, $E=M\pm 2\sigma_M$ three-point
scan appear in Fig.~\ref{fig7}.  For $\mpzero\sim 150\gev$ ($\sim 200\gev$)
and $\Delta\sigma_M/\sigma_M=0.01$, one finds systematic error
of $\Delta\gampzero/\gampzero\sim 0.028$ ($0.022$), which is
indeed an acceptable level. However, as described in the previous Section,
despite the factor of 2.2 increase [see Eq.~(\ref{lform})] in yearly
luminosity achieved by increasing $R$ from $0.003\%$ to $0.01\%$,
the decrease in the signal to background ratio is very substantial
and would lead to worse statistical errors unless $\mpzero>200\gev$
[for which $\gampzero/\sigm(R=0.003\%)>5$]. Thus, for $\mpzero<200\gev$
and typical model parameters as embodied in the choices of Ref.~\cite{mumu},
one should employ $R=0.003\%$ for the scan.

Let us now consider the $r_c$-ratio technique for the $P^0$. We will
compare to the scan technique using the choices $R=0.003\%$ for $\sigma_M^{\rm
min}$ and $R=0.03\%$ for $\sigma_M^{\rm max}$. This means $\sigma_M^{\rm
central}\sim 6.3\mev\,(M_{P^0}/100\gev)$. From Fig.~\ref{fig6}, we then
find $\gampzero/\sigma_M^{\rm central}\sim 2/3$ at $\mpzero=50\gev$
rising slowly to 1.6 at $\mpzero=200\gev$. 
This region is that for which the slope $|s|$ (see Fig.~\ref{dlgdlr}) 
is smallest. Consequently, the error in $\gampzero$
will be small if that for $r_c$ is. 
The $\Delta\sigma_c/\sigma_c$ errors for 3 years of operation at
pessimistic instantaneous luminosity ($L=0.3\fbi$) at $R=0.003\%$
were given in Fig.~19 of Ref.~\cite{mumu}. We rescale these
errors to $L=0.2f\fbi$ (corresponding to $2f$ years of operation
at $R=0.003\%$), where $f$ will be chosen to minimize the error in $r_c$. 
We also compute $\Delta\sigma_c/\sigma_c$ for $L=0.94(2-f)\fbi$
devoted to $R=0.03\%$ running (corresponding to $4-2f$ years
of operation at this latter $R$).
The net $\Delta\sigma_c/\sigma_c$ is computed using Eq.~(\ref{sigcerror})
after combining all final state channels.
We also compute $\Delta r_c/r_c$ according to the Eq.~(\ref{sigcerror})
procedure.  The corresponding statistical error for $\gampzero$ is then
computed using the appropriate $|s|$ slope value. We then search
for the value of $f$ (see above) such that $\Delta\gampzero/\gampzero$
is smallest. The value of $f$ and the corresponding errors 
for the combined-channel
$\sigma_c$ and for $\gampzero$ error are given in Table~\ref{fmcerrorspgb}
for the same $\mpzero$ values as considered for the three-point scan procedure.
For $\sigma_c$, the $r_c$-ratio procedure statistical errors
are very similar to the 4-year three-point scan statistical errors
for all $\mpzero$ values considered. For $\gampzero$,
the $r_c$-ratio procedure statistical errors are as good
as the tabulated 4-year three-point scan statistical errors
for $\mpzero<110\gev$, 
and become superior for larger $\mpzero$ values where $\gampzero/\sigm^{\rm
central}$ is significantly larger than unity.

As we have discussed, for the $r_c$-ratio procedure,
the fractional systematic error in $\gampzero$
is equal to that in $\sigma_M$ and there is no systematic
error in $B\equiv B_{\mupmum}B_F$. 
In contrast, we have seen that for $\mpzero<80\gev$
the systematic errors in $\gampzero$ and $B$ from
the scan technique will be somewhat larger than $\Delta\sigm/\sigm$.
Taking $\gampzero$ as an example, the scan systematic errors are of
order $1.5\Delta\sigm/\sigm$. As summarized earlier, this means that
the scan systematic error for $\Delta\sigm/\sigm=0.01$
is essentially the same as the
scan statistical error computed assuming pessimistic luminosity.
For optimistic luminosity the scan systematic error would be dominant.
Thus, the $r_c$-ratio procedure will actually give better overall
(systematic plus statistical) error for $\gampzero$ and, especially,
$B$ than the scan procedure for low as well as high $\mpzero$
values. This will become especially important if the luminosity available
is better than the pessimistic value or if $\Delta\sigm/\sigm>0.01$.
If a narrow resonance is observed in the $\gam\gam$ final state at
the Tevatron or LHC, and if it has the weak coupling
to $ZZ$ that is typical of a pseudogoldstone boson,
then the $r_c$-ratio technique for precision measurements of its properties
at the muon collider is strongly recommended.

\bigskip

\begin{figure}[p]
\epsfysize=16truecm
\centerline{\epsffile{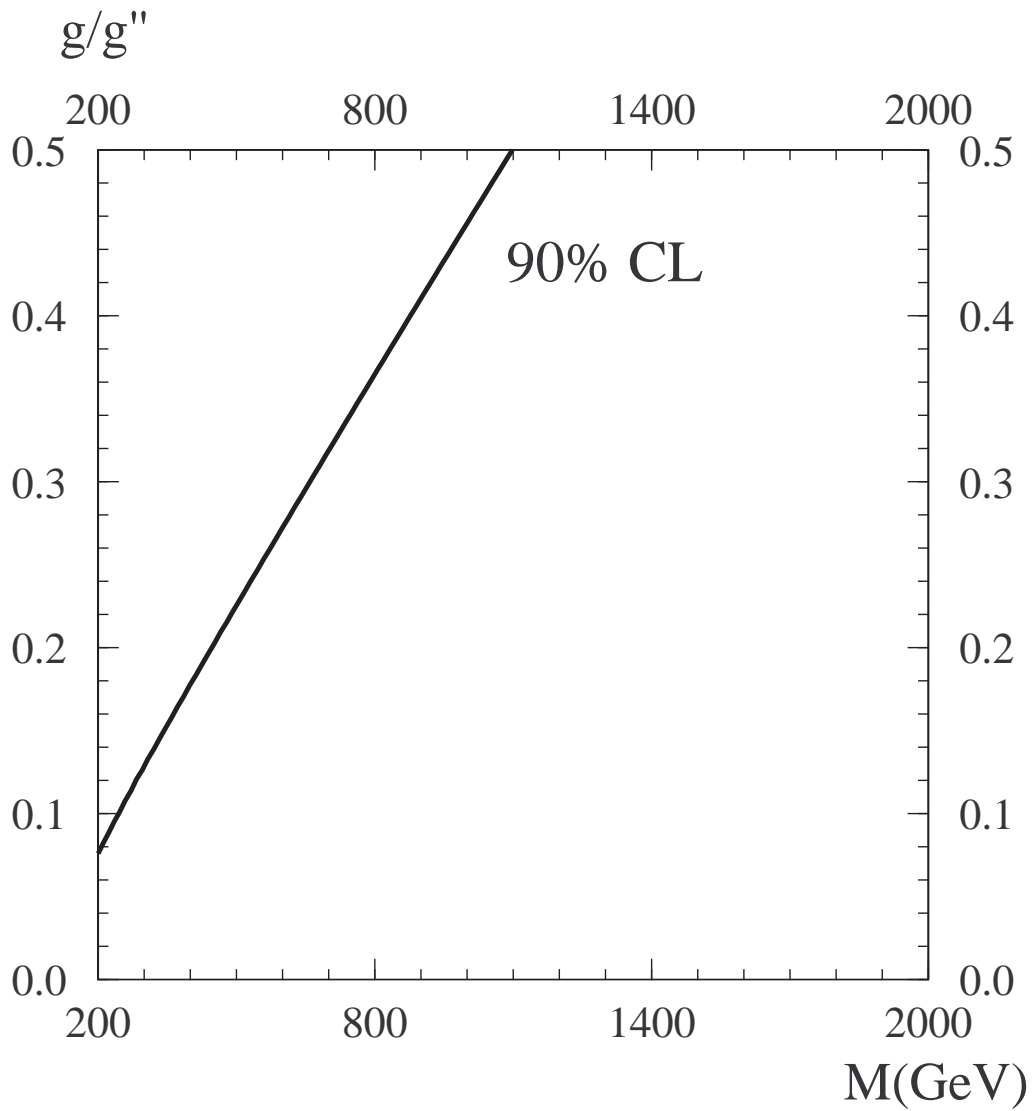}}
\noindent
\caption{
The bound in the $(M,g/g'')$ parameter plane of the Degenerate BESS
model arising from existing precision experimental data. The region
to the left of the solid line is disallowed at $\geq 90\%$ CL.}
\label{fig8}
\end{figure}

\begin{figure}[p]
\epsfysize=16truecm
\centerline{\epsffile{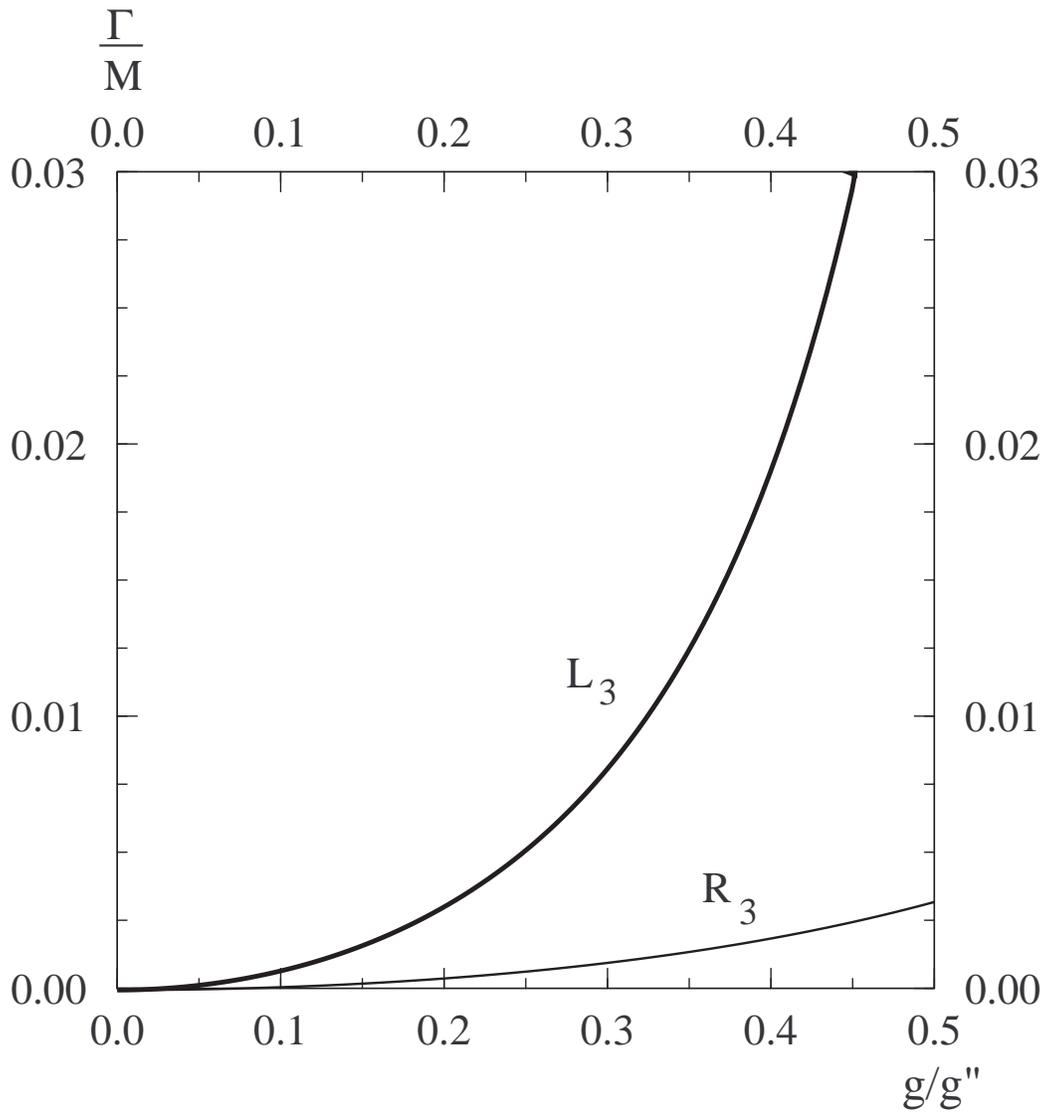}}
\noindent
\caption{
The ratio $\Gamma/M$ as a function of $g/g''$ for the
Degenerate BESS for the neutral vector resonances $L_3$ and
$R_3$.}
\label{fig9}
\end{figure}

\begin{figure}[p]
\epsfysize=16truecm
\centerline{\epsffile{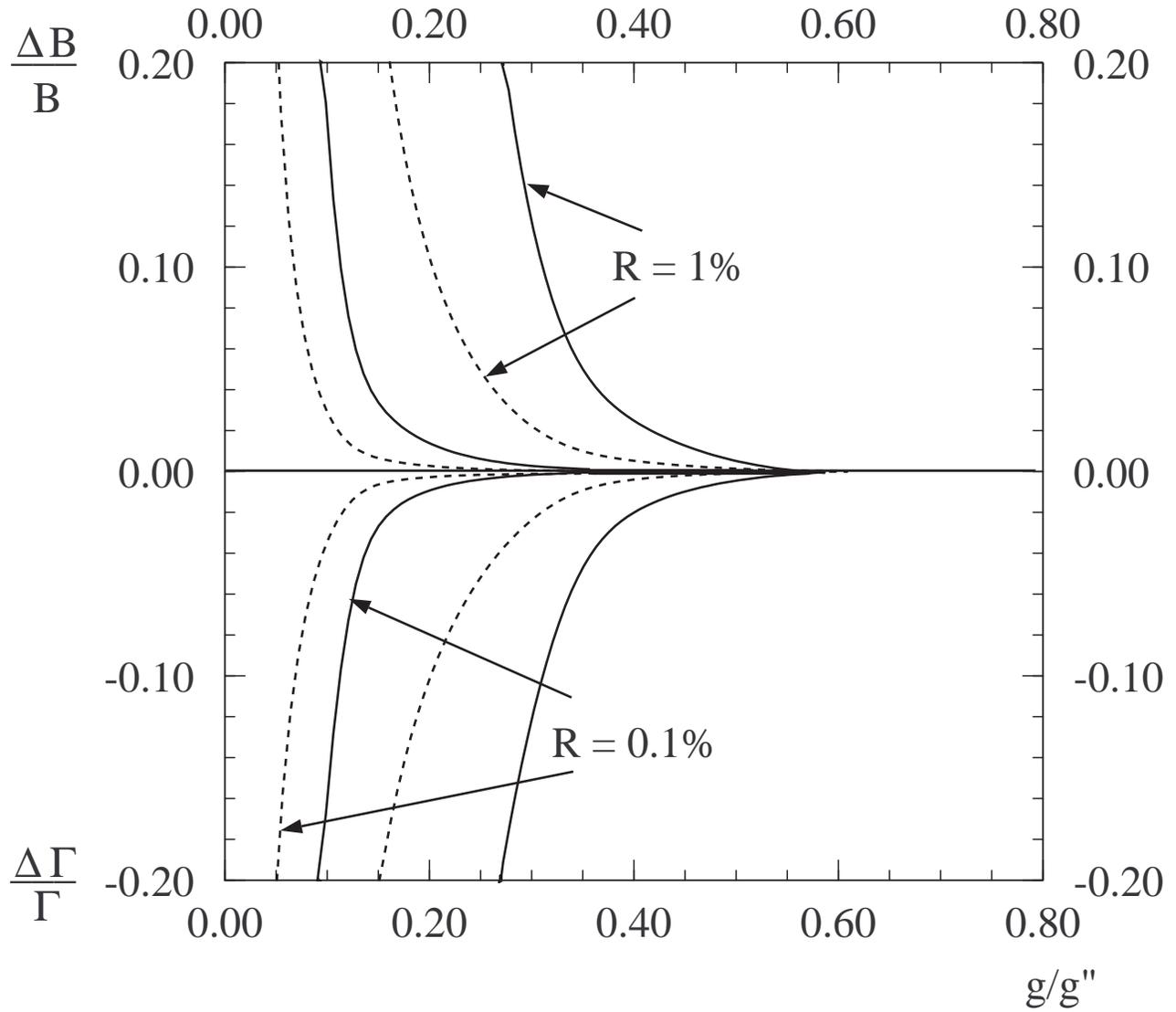}}
\noindent
\caption{
The Degenerate BESS model: the fractional errors on $B$
and $\Gamma$ for the vector resonance $L_3$ as functions of
$g/g''$ from the three-point scan measurement ($E=M$, $E=M\pm
2\Gamma$), for $\Delta\sigma_M/\sigma_M=0.05$ (solid line) and
$\Delta\sigma_M/\sigma_M=0.01$ (dashed line), and for $R= 0.1\%$
and $R=1\%$.}
\label{fig10}
\end{figure}

\begin{figure}[p]
\epsfysize=16truecm
\centerline{\epsffile{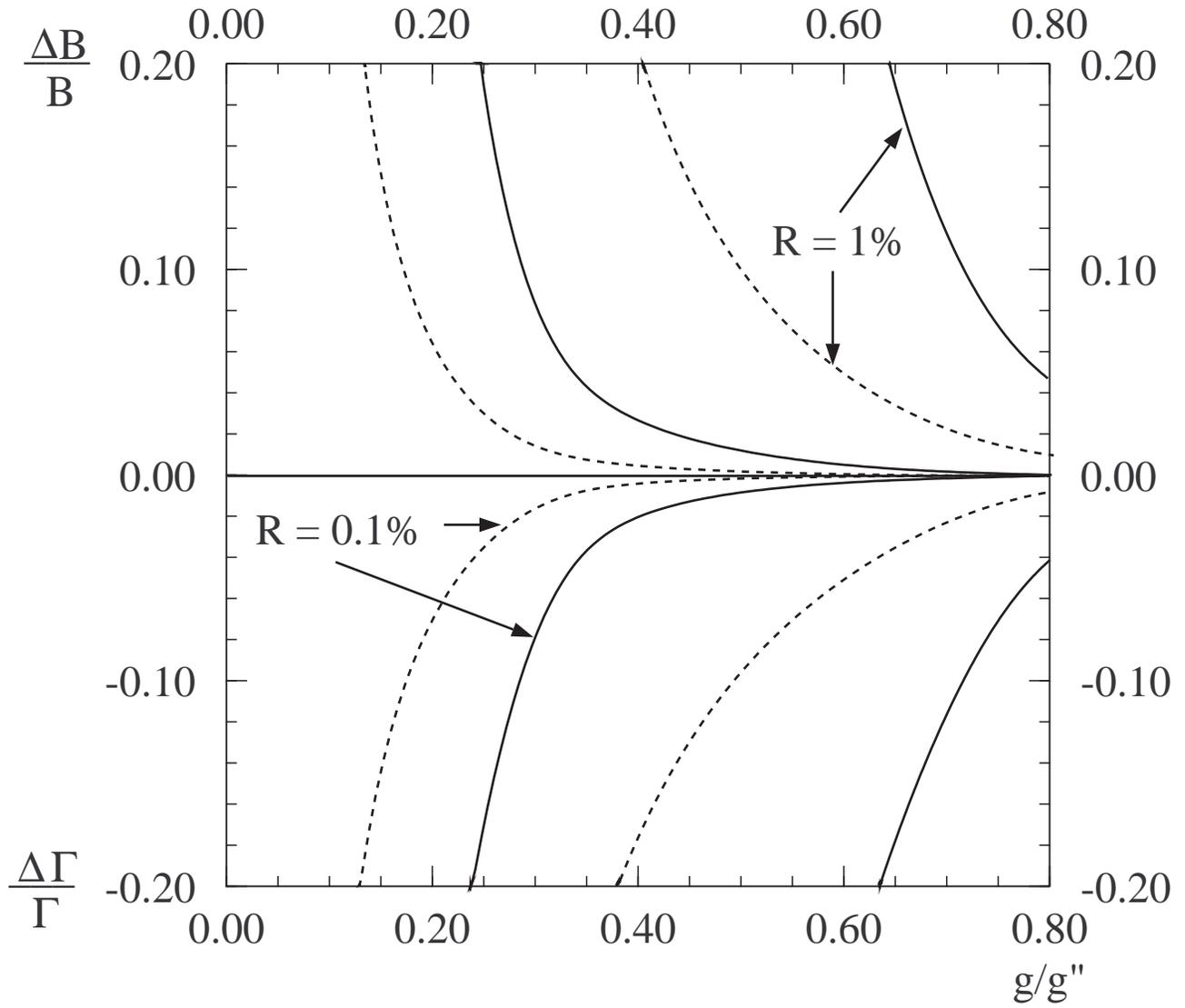}}
\noindent
\caption{
Same as in Fig. \ref{fig10} for the vector resonance $R_3$.}
\label{fig11}
\end{figure}

\noindent {\bf{Degenerate BESS}} -
The Degenerate BESS model \cite{DBESS} describes
two isotriplets
of nearly degenerate vector resonances $(\vec L,\vec R)$
characterized by two parameters $(M,g'')$, the common mass (when
the EW interactions are turned off) and their gauge coupling. The
main feature of the model is the decoupling property, which
implies very loose bounds from existing precision experiments
\cite{vancouver} as shown in Fig. \ref{fig8}. Another consequence is that
the decay of the resonances into pairs of ordinary gauge vector
bosons is quite depressed. The total widths for the two neutral
states $(L_3,R_3)$ are given by
\be
\Gamma_{L_3,R_3}=M\,h_{L_3,R_3}(g/g'')\,,
\ee
where $g$ is the weak coupling constant. The behaviour of the
total widths as functions of $g/g''$ is
shown in Fig. \ref{fig9}, whereas for $g/g''\ll
1$ one has
\be
h_{L_3}\approx 0.068\left(\frac g{g''}\right)^2,~~~ h_{R_3}\approx
0.01\left(\frac g{g''}\right)^2\,.
\ee
The ratio of the widths, $\Gamma_{L_3}/\Gamma_{R_3}$, in the
interval $0\le g/g''\le .5$ approximately varies between $0.15$
and $0.07$. The weak interactions break the mass degeneracy
giving rise to the mass splitting
\be
M^s=M_{L_3}-M_{R_3}=M\,h_{M^s}(g/g'')\,.
\ee
The behaviour of $h_{M^s}$ for  $g/g''\ll 1$ is
\be
h_{M^s}\approx (1-\tan^2\theta_W)\left(\frac
g{g''}\right)^2\approx 0.7 \left(\frac g{g''}\right)^2\,.
\ee
The branching ratios into charged lepton pairs are almost parameter
independent and rather sizeable
\be
B(L_3\to\ell^+\ell^-)\approx
4.5\%,~~~~B(R_3\to\ell^+\ell^-)\approx 13.6\%\,.
\ee
In this Section, we will assume that $\sigma_M$ is much smaller than
$M^s$, and therefore we can apply the previous analysis to
the two resonances separately. The case $\sigma_M\approx
M^s\gg
\Gamma_{L_3,R_3}$ will be studied in Section 4.
By combining Fig. \ref{fig2} with Fig. \ref{fig9}, 
we easily obtain the $\Delta\sigma_M$-induced
fractional errors in the branching ratios and in
the widths of the two resonances $L_3$ and $R_3$ as functions of
$g/g''$ for different choices of $\sigma_M$ [equivalently, $R$; see Eq.
(\ref{spread})]. The results are given in Figs. \ref{fig10} and \ref{fig11} for
$L_3$ and $R_3$, respectively,  for the choices
$\Delta\sigma_M/\sigma_M=1\%$ and 5\%, and for $R=1\%$
and 0.1\%. (Because the resonance widths are large,
we do not need the very small values of $R$ required
to study the SM Higgs or the $P^0$.)
Combining these results with the bounds on the
portion of parameter space still allowed
by precision experiments, one can put lower limits on the masses of
the resonances such that they have not been excluded and yet
one is able to measure the widths with no more than a
given systematic uncertainty from $\Delta\sigma_M$.
For instance, in the case of a machine with $R=1\%$ (typical of an
$e^+e^-$ collider) and for $\Delta\sigma_M/\sigma_M=1\%$,
one finds, from Fig. \ref{fig9}, that
$g/g''\gsim 0.3$ is needed to avoid $\Delta\sigma_M$-induced
errors in $\Gamma_{L_3}$ above $2\%$. From Fig. \ref{fig8}, the
portion of parameter space that is still allowed by precision
experiment can be roughly expressed by
\be
\frac{M}{1000\gev}\gsim 2.17\left(\frac g{g''}+0.01\right)\,.
\ee
Therefore, the $g/g''\gsim 0.3$ requirement converts to the requirement
that $M\gsim 670\gev$.
This means that, at lepton colliders with $R=1\%$ and
$\Delta\sigma_M/\sigma_M=1\%$, one will be able to measure
$\Gamma_{L_3}$ (for an allowed $L_3$ resonance)
with a $\Delta\sigma_M$-induced error
of less than 2\% only if the mass
of the resonance is greater than 670 GeV. Similar
considerations apply to $R_3$, where keeping the systematic
error in $\Gamma_{R_3}$ below $2\%$ requires
$g/g''\gsim 0.7$, leading to $M\gsim 1540$ GeV.
In the case of a muon collider, $R=0.1\%$ can be achieved
while maintaining large luminosity. In this case,
the $\Delta\sigma_M$-induced fractional error for $L_3$
is below $2\%$, for $\Delta\sigma_M/\sigma_M=1\%$, if $g/g''\gsim 0.12$,
which converts to $M\gsim 280\gev$ for $L_3$
resonances not already excluded by the precision data. The corresponding
limits for $R_3$ are $g/g''\gsim 0.3$
and $M\gsim 670\gev$. In particular, we see from Fig. \ref{fig8} that,
at a machine working at the top threshold (about 350 GeV), only
resonances with $g/g''\lsim 0.14$ have not already been
excluded by the precision data. Therefore,
only a machine with $R\le 0.1\%$ would be able to measure the
width and branching ratios of an allowed resonance without
encountering significant systematic uncertainty
coming from $\Delta\sigma_M/\sigma_M\sim 0.01$.

\bigskip

\section{Analysis for nearly degenerate resonances}

\bigskip
In this Section we will discuss  nearly degenerate
resonances, \ie\ the case of a mass splitting much smaller
than the average mass of the resonances, $M^s=M_2-M_1\ll M=(M_1+M_2)/2$.
We will be interested in the case where the energy spread
of the beam is of the same order of magnitude as the mass
splitting between the two peaks, $\sigma_M\approx M^s$. We
will  also assume that the widths of the two resonances are much
smaller than the mass splitting, \ie\ $\Gamma_1,\,\Gamma_2\ll
M^s$. It follows also that $\Gamma_1,\,\Gamma_2\ll\sigma_M$. In this
approximation, we can safely describe the cross section as the sum
of two Breit-Wigner functions, and furthermore we can use the
narrow width approximation. Therefore,~\footnote{We
focus on the total cross section, but it should be kept in mind
that we could also consider the cross section in a given final state.
In this case, $\Gamma(R_i\to\ell^+\ell^-)$ ($i=1,2$) should be replaced by
$\Gamma(R_i\to\ell^+\ell^-)B_F$ in all that follows.
The $\Delta\sigma_M$-induced systematic errors on this product
would be the same as for $\Gamma(R_i\to\ell^+\ell^-)$.}
\begin{equation}\label{bw2}
M^2\sigma_c=
B_{\ell^+\ell^-}^1\Phi(x_1,\gamma_1)+B_{\ell^+\ell^-}^2\Phi(x_2,\gamma_2)\,,
\end{equation}
where $B_{\ell^+\ell^-}^1$ and $B_{\ell^+\ell^-}^2$ are the branching
ratios of the  resonances into $\ell^+\ell^-$, and
\begin{equation}\label{def2}
x_1=\frac{E-M_1}{\sigma_M},~~~x_2=\frac{E-M_1-M^s}{\sigma_M}=x_1-m^s
\end{equation}
and
\begin{equation}\label{def3}
M_2=M_1+M^s,~~~m^s=\frac{M^s}{\sigma_M}\,.
\end{equation}
We have also assumed $M\approx M_1\approx M_2$. For a Gaussian
beam we recall that the function $\Phi(x,\gamma)$ defined in Eq.
(\ref{scaled-conv})  is given by
\begin{equation}\label{Phi}
\Phi(x,\gamma)=\sqrt{\frac \pi 2} (2 j+1)\int_{-\infty}^{+\infty}e^{-(x-y)^2/2}
\frac{\gamma^2}{y^2+\gamma^2/4}dy\,.
\end{equation}
We may evaluate this expression by
performing the Fourier transforms of the Gaussian distribution and
of the Breit-Wigner and then taking the inverse Fourier
transform. We get
\begin{equation}\label{integration}
 \Phi(x,\gamma)= (2j+1)\pi^{3/2}\gamma\int_{-\infty}^{+\infty}dp
 e^{-p^2/2+ipx}
\left(\theta(p)e^{-p\gamma/2}+
\theta(-p)e^{+p\gamma/2}\right)\,.
\end{equation}
Since we are assuming that the resonance widths are much
smaller than the energy spread, we may evaluate Eq.~(\ref{integration})
in the $\gamma\to 0$ approximation, yielding
\begin{equation}\label{approx}
\Phi(x,\gamma)=(2j+1)\gamma\sqrt{2}\pi^{3/2} e^{-x^2/2}\,,
\end{equation}
as earlier given in Eq.~(\ref{smallgam}) for $j=0$. As previously noted,
this expression shows that when the energy spread is much bigger than
the width the convolution gives rise to a Gaussian function
with spread $\sigma_M$, and we loose any information
about the total width. In fact, the total cross section $\sigma_c$
depends only on the product
$B_{\ell^+\ell^-}\gamma=\Gamma(R\to\ell^+\ell^-)/\sigma_M$.
In the case of two resonances with $\gamma_1,\,\gamma_2\to 0$, we
thus find
\begin{equation}\label{final cross}
M^2\sigma_c=(2j+1)\sqrt{2}\pi^{3/2}\left( g_1e^{-x_1^2/2}+g_2
e^{-(x_1-m^s)^2/2}\right)\,,
\end{equation}
where
\be
g_i=\gamma_iB^i_{\ell^+\ell^-}=\frac{\Gamma(R_i\to\ell^+\ell^-)}{\sigma_M}\,.
\ee
The function (\ref{final cross}) is invariant under the substitution
\be
g_1\leftrightarrow g_2,~~~x_1\leftrightarrow m^s-x_1
\label{invariance}
\ee
The behaviour of the function is characterized by the ratio
\be
\frac{g_2}{g_1}=\frac{\Gamma(R_2\to\ell^+\ell^-)}{\Gamma(R_1\to\ell^+\ell^-)}
\equiv
a
\ee
and by $m^s$. For small $m^s$, the convolution of the Gaussian with the
two Breit-Wigners has a single maximum in between 0 and $m^s$
depending on the value of $a$. For instance, for $a=1$ the
maximum is at $m^s/2$. In this situation, the second derivative
of the function (\ref{final cross}) has two zeros corresponding
to the changes of curvature before and after the
peak. By increasing $m^s$ the second derivative acquires a third
zero (in fact, a double zero). This is due to the effect of the smaller
Breit-Wigner which gives rise to a further change of the
curvature. Just to get an idea, we list in Table 5, for several choices of
$a$, the critical value $m^s_1$ of $m^s$ at which this
third zero occurs.
As $m^s$ is increased further, there comes a point
at which the two Breit-Wigner maxima start to
show up. The minimum value of $m^s$ required
to see two maxima,  $m^s_2$,
is given as a function of the ratio $a$ in Table \ref{doublezero}.
Notice that the invariance (\ref{invariance}) implies $m^s_i(a)=m^s_i(1/a)$.
\begin{table}[bht]
\caption{Values of $m^s$ at which the double
zero of the second derivative of the function (\ref{final cross})
occurs ($m^s_1(a)$) and at which the two maxima of the function
(\ref{final cross}) start to show up ($m^s_2(a)$).}
\begin{center}
\begin{tabular}{|c|c|c|}
\hline
$a$ & $m^s_1(a)$ & $m^s_2(a)$ \\
\hline\hline
1 & 2 &2\\
\hline
2 & 1.85 & 2.63 \\
\hline
3 & 2.02 & 2.85 \\
\hline
4 & 2.20 & 2.98\\
\hline
5 & 2.31 & 3.08\\
\hline
\end{tabular}
\end{center}
\label{doublezero}
\end{table}

We  can now  discuss the type of measurements necessary to
determine the parameters $M_1$, $M_2$, $g_1$ and $g_2$.
In this discussion, we will use the notation $x\equiv x_1$.
We first determine the overall location in energy of the
resonance structure by locating the absolute maximum
of the cross section.
For small $m^s$, such that the individual resonance peaks
are unresolved, the cross section has a single maximum near
the location of the resonance with the larger $g$; we will assume
that it is $g_2$ which is largest.  For $m^s$ large enough
that the peaks are resolved, we may locate the larger maximum.
We denote the location of the maximum in the cross section by $E_{\rm max}$,
and choose this as our first energy setting. We write
\be
E_{\rm max}=\sigma_{M} x_{\rm max} +M_1\,,
\label{energyscale}
\ee
where $x_{\rm max}$ is a function of $a$ and $m^s$ which can be
evaluated numerically from Eq. (\ref{final cross}). For instance,
for $a\gsim 2$ it is a good approximation to assume $x_{\rm
max}\approx m^s$ (specially for $m^s$ bigger than the critical
value $m^s_2$), implying $E_{\rm max}\simeq M_2$.
The value of the cross section at $x_{\rm max}$ provides a second
input into determining $M_{1,2}$ and $g_{1,2}$.
To complete the process of determining these four parameters,
we need two more measurements.  If $m^s$ is larger than $m_2^s$,
so that a second (lower) maximum is present, we can use the
location of the second maximum and the cross section value
at this second maximum as our two additional measurements.
We effectively have four equations in four unknowns, two equations
involving the derivatives of Eq.~(\ref{final cross}) and two involving
the absolute magnitude of Eq.~(\ref{final cross}).
If no second maximum is present, then we must effectively determine
the slope of $\sigma_c(x)$ of Eq.~(\ref{final cross}) at some energy
location away from $x_{\rm max}$ and determine
the cross section at this same location. Measurements of $\sigma_c(x)$
at two nearby values of $x$ away from $x_{\rm max}$ are needed.
That these approaches are really
equivalent becomes apparent when we realize that the $m^s>m_2^s$ procedure
of locating the second maximum actually requires measuring the slope
of $\sigma_c(x)$ and finding its second zero.
(Recall that this discussion assumes infinite statistics so that
we will end up effectively computing the minimum error that
will be induced by uncertainty in $\sigma_M$.)

In the absence of a second maximum, the choice of the
other two energies can be difficult if $m^s$ is smaller than the
critical value $m^s_1$. In fact, in this case the
convolution of the two Breit-Wigner looks very much like the
convolution with a single Breit-Wigner. Not surprisingly,
to obtain reasonable errors
it is necessary that $\sigma_M$ be such that $m^s$ is at least bigger than
$m^s_1$. Let us assume that for $m^s>m^s_1$ we can
approximately locate the energy corresponding to $x\sim 0$ and that we
measure the cross section at two points
in its vicinity (as well as at $x=x_{\rm max}$). If
$m^s$ is greater than  $m^s_2$, we may continue to employ the
above procedure or we may use the alternative procedure
outlined earlier based on the fact that the
cross section has two maxima, one near $x=0$ and one near $x=m^s$;
in the alternative procedure, we measure the energy and
the cross section at the two peaks.

\begin{figure}[p]
\epsfysize=16truecm
\centerline{\epsffile{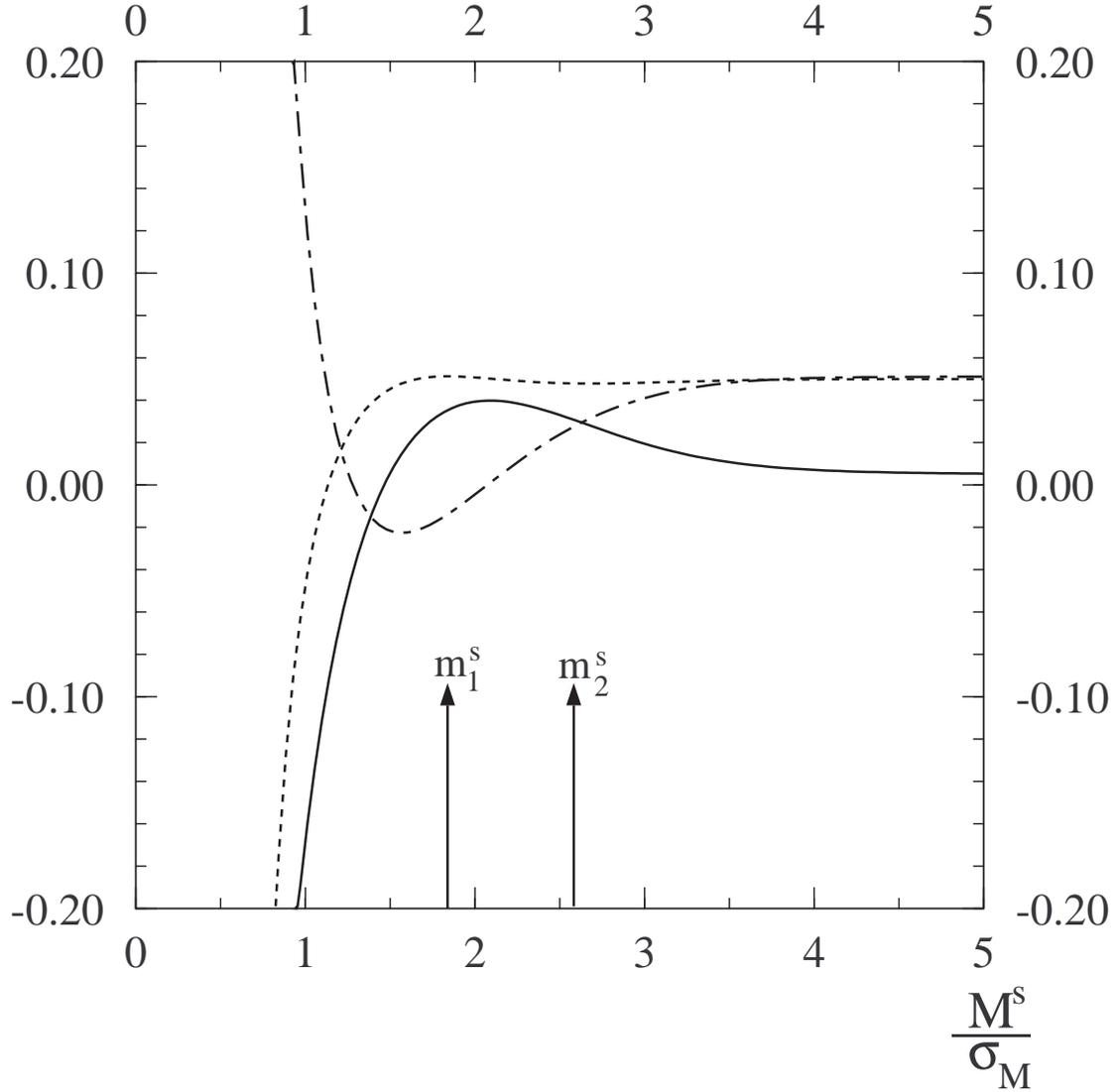}}
\noindent
\caption{
The fractional errors in the partial widths, $\Delta
\Gamma(R_i\to\ell^+\ell^-)/\Gamma(R_i\to\ell^+\ell^-)$, $i=1,2$
(dashed and dot-dashed lines respectively), and in the mass
splitting, $\Delta M^s/M^s$ (continuous line), induced
by an energy spread uncertainty of $\Delta\sigma_M/\sigma_M=0.05$,
as a function of $m^s$ for
$a=\Gamma(R_2\to\ell^+\ell^-)/\Gamma(R_1\to\ell^+\ell^-)=2$. The
input cross section measurements are at
$x_{max}$, see Eq.~(\ref{energyscale}), $x=0.1$ and $x=0.2$.}
\label{fig12}
\end{figure}

\begin{figure}[p]
\epsfysize=16truecm
\centerline{\epsffile{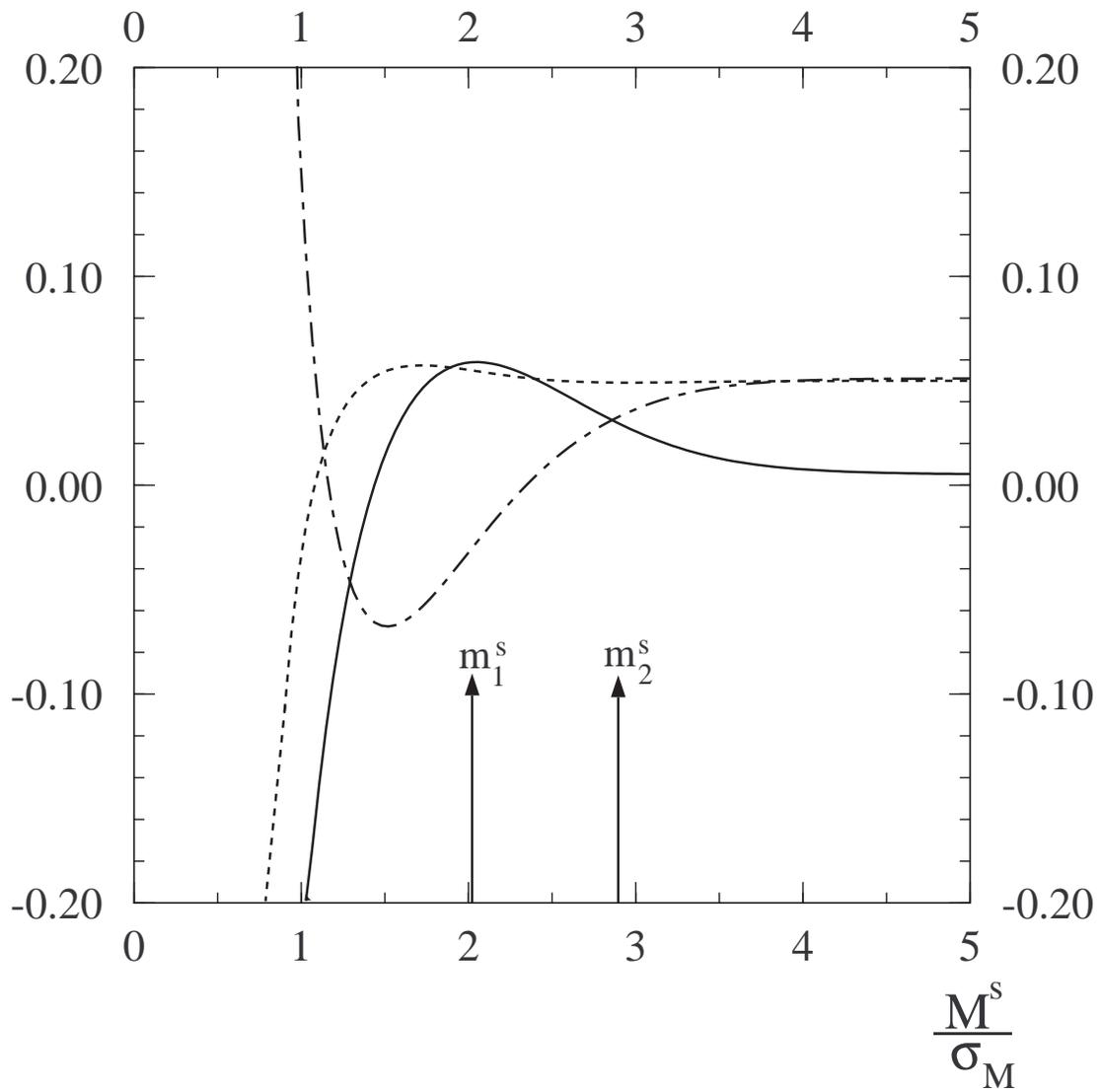}}
\noindent
\caption{
Same as in Fig. \ref{fig12}, with $a=3$.}
\label{fig13}
\end{figure}

We consider first the former procedure that is the only choice if $m^s<m^s_2$.
We start our analysis by taking $a=2$ and then later discuss
the modifications for different values of $a$. From the
measurement which fixes the energy scale (see Eq. (\ref{energyscale}))
we find the following relation between the parameter errors
and the uncertainty in the energy spread:
\be
\left(1+\frac{\Delta\sigma_M}{\sigma_M}\right)x_{\rm max}(m^s+\Delta m^s)
+\Delta m_1-x_{\max}(m^s)=0\,,
\label{energy1}
\ee
where $\Delta M_1=\sigma_M\Delta m_1$ is the error in $M_1$ and
$\Delta m^s$ is the error in $m^s$. From this equation we can
eliminate $\Delta m_1$ in terms of the other
errors. This has been done by using the approximation $x_{\rm
max}\approx m^s$. The errors on $M^s$ and on the partial
widths can then be determined using the three
cross sections --- $\sigma_c(x_{\rm max})$
and $\sigma_c(x)$ at two other $x$ values near 0 ---
following a procedure analogous to that discussed for
a single resonance using Eq. (\ref{condition1}).
In particular, we assume measurements of the
cross section at  $x_{\max}$ and at $x=0.1$ and $x=0.2$. The
resulting fractional errors for $\Gamma(R_i\to\ell^+\ell^-)$ ($i=1,2$)
and $M^s$ are given in Fig. \ref{fig12} as a function of $M^s/\sigma_M$
for $\Delta\sigma_M/\sigma_M=5\%$. As
expected, the errors grow rapidly once $m^s$ falls below $m^s_1$.
As $m^s$ is increased above $m^s_1$, there is a change of curvature
in the fractional error curves  around the critical value $m^s_2$,
after which the fractional errors approach asymptotic limits.
Notice that the asymptotic value of $\Delta M^s/M^s$ is zero,
whereas $\Delta\Gamma(R_i\to\ell^+\ell^-)/\Gamma(R_i\to\ell^+\ell^-)\to
\Delta\sigma_M/\sigma_M$ for both $i=1,2$. This is because the cross sections
depend only on the ratio $\Gamma(R_i\to\ell^+\ell^-)/\sigma_M$.
In Fig. \ref{fig13} we represent the same quantities but for $a=3$, and we
see results very similar to those for $a=2$ except
for changes due to the different values of the critical
points $m^s_1$ and $m^s_2$. We have tried different choices for the
two measured points off the maximum, varying them up to
$x=0.3$, without any significant change in the results.

\begin{figure}[p]
\epsfysize=16truecm
\centerline{\epsffile{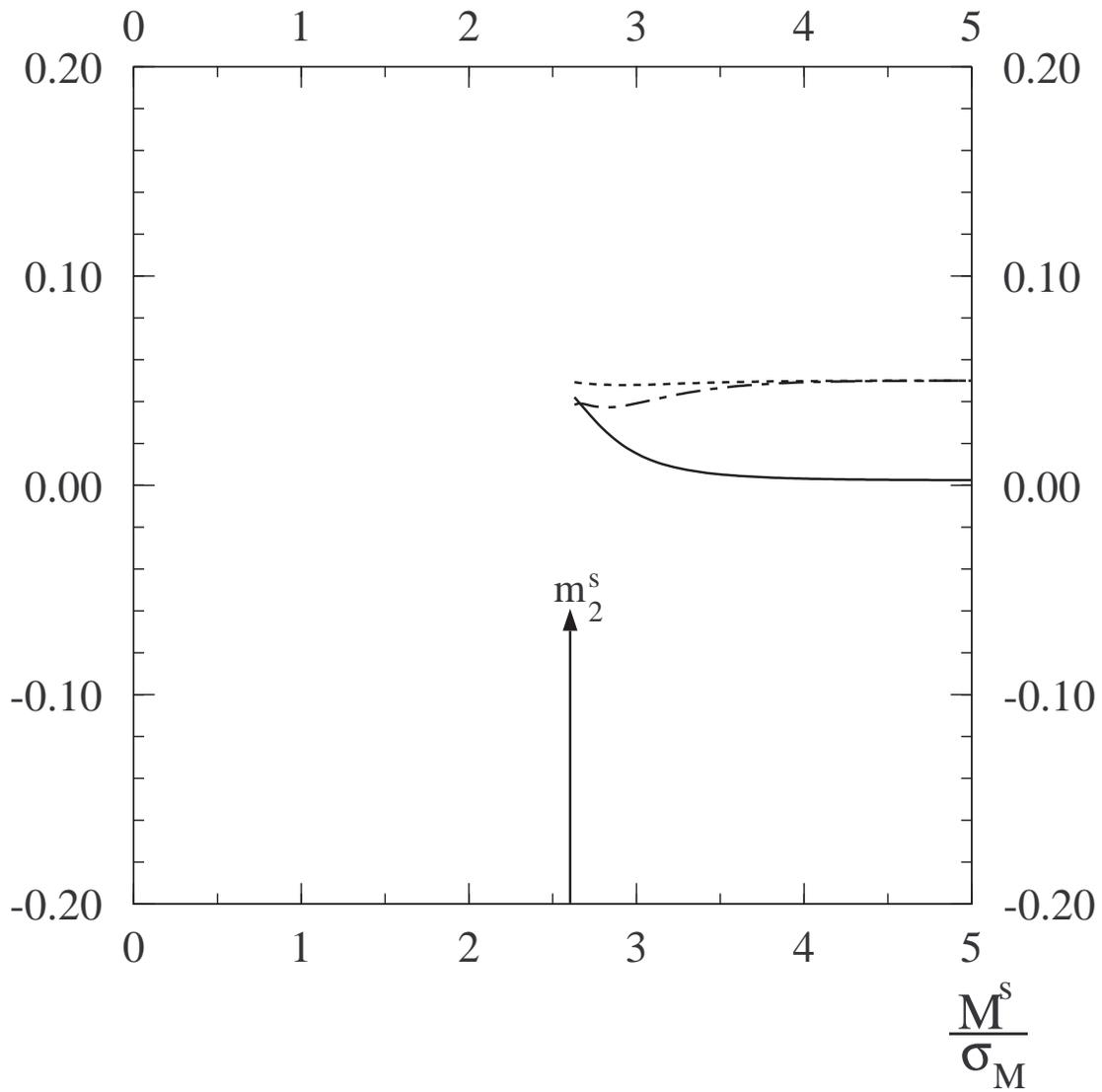}}
\noindent
\caption{
Same as in Fig. \ref{fig12}, again with $a=2$ but using the positions
of the two maxima and the cross section values at the two maxima for
determining the fractional errors.}
\label{fig14}
\end{figure}

We now consider the alternative procedure outlined earlier that
is possible when two cross section maxima become visible,
that is when the mass splitting is
bigger than  $m^s_2$. In this case, the four
measurements for determining $M_{1,2}$ and $g_{1,2}$
are the energy locations of the two maxima and the
cross sections at these two maxima. The measurement of the energy of the
second maximum gives rise to a condition similar to the one of Eq.
(\ref{energy1}). The resulting fractional errors
in $M^s$ and $\Gamma(R_i\to\ell^+\ell^-)$ are given in Fig. \ref{fig14} 
(for $a=2$), and are essentially the same as obtained
in the previous procedure when $m^s>m^s_2$.

The basic conclusion from these analyses is that for $m^s$ of
the order of the critical value $m^s_2$, the fractional
errors in the parameters of the resonances are of the order of
the fractional error in $\sigma_M$. For smaller $m^s$ the errors
become very large. For $m^s$ significantly bigger
than $m^s_2$, the fractional error in the
mass splitting rapidly approaches zero while the fractional errors for
the $\Gamma(R_i\to\ell^+\ell^-)$ partial
widths become equal to the fractional error in the energy spread.
In short, we can use the critical value $m^s_2$
in order to discriminate between a good and a bad determination
of the mass splitting.

\bigskip
\section{Application to Degenerate BESS}
\begin{figure}[p]
\epsfysize=16truecm
\centerline{\epsffile{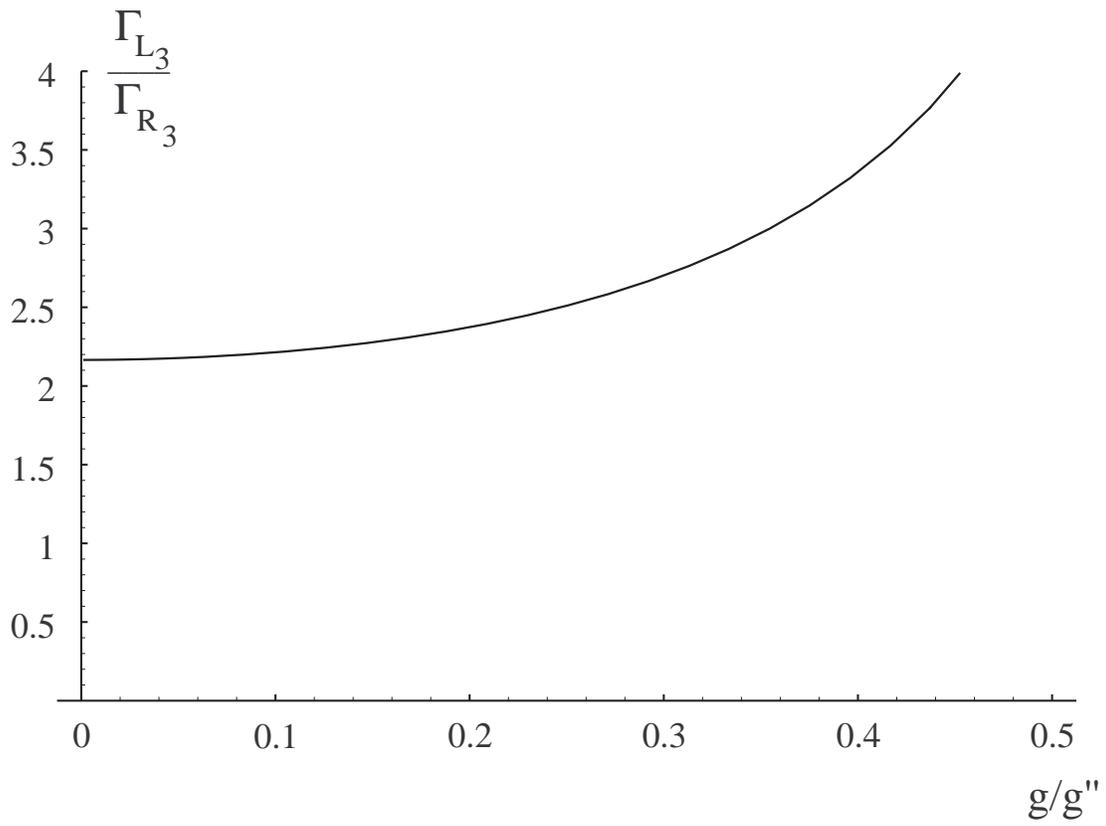}}
\noindent
\caption{
The ratio $\Gamma(L_3\to\ell^+\ell^-)/\Gamma(R_3\to\ell^+\ell^-)$
vs. $g/g''$ in the Degenerate BESS model.}
\label{fig15}
\end{figure}

 \bigskip
In Degenerate BESS one can show
that the condition $M^s
\gg\Gamma_L,\,\Gamma_R$ is rather well satisfied (by one and two
orders of magnitude respectively). Therefore we can apply the
analysis of the previous Section. From Fig. \ref{fig15}, we see that the
value of $a$ is almost constant and approximately 2.2 for $g/g''$
up to 0.2, and then $a$ increases up to $\approx 4$ for $g/g''=0.5$.
As discussed in the last Section, we can use the values of $m^s_2$
given in Table \ref{doublezero} in order to determine
the minimum value of $M^s/\sigma_M$ needed
in order to make a good determination of the mass
splitting. For $a=2.2$ one finds that the minimum
value is $m^s_2=2.68$. As in our earlier single resonance discussions,
for any fixed value of the energy resolution $R$, we
convert this bound into lower bounds for $g/g''$ and for the mass
$M$ of nearly degenerate resonances that have not already been
excluded by precision experimental data. From
$M^s/\sigma_M\gsim 2.68$ we get
\bea
R=1\% &\to& \frac g{g''}\gsim 0.16,~~~ M\gsim 370\gev\,,\nn\\
R=0.1\% &\to& \frac g{g''}\gsim 0.05,~~~ M\gsim 130\gev\,.\nn
\eea
We see that for $R=1\%$ a machine with energy near the top threshold would
just be at the border of being able to accurately measure
the mass difference between the $L_3$ and $R_3$ resonances not already
excluded by precision data.

\bigskip
\section{Conclusions}

We have considered the production of a narrow resonance
via $s$-channel collisions of leptons ($\ell=e$ or $\mu$).
Here, a `narrow' resonance is defined as one that has
width $\Gamma$ substantially smaller than the beam energy spread $\Delta E_{\rm
beam}$ that is natural for the collider (and therefore is associated with 
the largest instantaneous luminosity). It will be convenient to use
the parameterization $\Delta E_{\rm beam}=0.01R\,E_{\rm beam}$, 
where $R$ is in per cent. For example, at a muon collider
with center of mass energy $E\sim 100\gev$,
$R\sim 0.12\%$ allows for maximal $L$
and $L$ declines rapidly as $R$ is forced
to smaller values by compression techniques. A resonance with width
$\Gamma\ll 0.001M$ would then be narrow. Our focus has been on the
systematic error that might be introduced into measurements of the
parameters of a narrow resonance due to systematic uncertainty in the value of
$\Delta E_{\rm beam}/E_{\rm beam}$.
The important parameters that can be measured are the branching 
ratio of the resonance
into the charged leptons (\ie\ those that are being collided), the product
of this leptonic branching ratio times that for
the resonance to decay to a particular final state,
and the total width of the resonance. We examined four methods
for determining the resonance parameters: (1) a scan of the resonance;
(2) sitting on the resonance and changing the beam energy resolution;
(3) measurement of the cross section in the $\ell^+\ell^-$
final state; and (4) measurement of the Breit-Wigner area. 
Methods (3) and (4) avoid the introduction of systematic errors
due to uncertainty in $\Delta E_{\rm beam}/E_{\rm beam}$,
but for the integrated luminosities that are anticipated to be available
the statistical errors associated with these techniques
would be quite large for a narrow resonance. Methods
(1) and (2) can provide resonance parameter determinations with small
statistical error. However, even in the limit of infinite statistical accuracy, 
determinations of the resonance parameters
are sensitive to systematic uncertainties in $\Delta E_{\rm beam}$
if $\Gamma$ is not much larger than $\Delta E_{\rm beam}$.
At a muon collider, the smallest $R$ that can be achieved is
expected to be $R=0.003\%$, for which $\Delta E_{\rm beam}\sim \Gamma_H$ 
for a light SM Higgs boson and $\Delta E_{\rm beam}\sim \gampzero/2$
for the lightest pseudo goldstone boson of a technicolor model.
Consequently, in these and other similar cases, a detailed assessment of 
the systematic errors in resonance parameter determinations 
introduced by uncertainty in $\Delta E_{\rm beam}$ is very important.

We have performed a general analysis to determine the (systematic) errors
in the measured resonance parameters induced by a systematic uncertainty
in $\Delta E_{\rm beam}$. We find that the
induced fractional errors in the leptonic branching ratio
(and also the product of the leptonic branching
ratio times the branching ratio into any given final state)
and in the total width can be expressed as universal
functions of the ratio $\Gamma/\sigma_M$, where $\sigma_M$
is the spread in total center of mass energy resulting
from the beam energy spreads: $\sigma_M/M=0.01R/\sqrt 2$.
In the case of the minimal three-point scan, with sampling
at $E=M,E=M\pm k\Gamma$, the error functions also depend on $k$.

For a minimal three-point scan with
$k=1$, the induced $\Delta\Gamma/\Gamma$ fractional systematic
errors were parameterized as a function of $\Gamma/\sigma_M$
in Eq.~(\ref{dgamform}). Very roughly,
for $\Gamma/\sigma_M\sim 2.5$ we find that the
induced fractional errors in $\Gamma$ and $B=B_{\ell^+\ell^-}$
or $B_{\ell^+\ell^-}B_F$ ($F$=final state) are of the order of the
fractional uncertainty $\Delta\sigma_M/\sigma_M$.
As $\Gamma/\sigma_M$ increases above 2, the fractional errors smoothly
decrease. For values of $\Gamma/\sigma_M$ below 1, the
fractional errors in the resonance parameters increase very rapidly.
For example, for $\Gamma/\sigma_M\sim 1$ ($\sim 0.2$)
the $\Delta\sigma_M$-induced
fractional systematic errors in the resonance parameters increase to
$\sim 3.5\div 4\Delta\sigma_M/\sigma_M$ 
($\sim 20\div 25\Delta\sigma_M/\sigma_M$),
for the $k=1$ scan. Thus, to avoid large systematic errors from
$\Delta\sigma_M$, it is imperative to operate the collider with
$\Gamma/\sigma_M$ no smaller than 1. If $R$ can be adjusted to achieve
values significantly larger than 1, one can consider how to optimize
the choice of $R$ so as to minimize the total statistical plus
systematic error.  A discussion was presented leading to the
following two basic conclusions.
(a)
For a broad resonance, defined as one
with $\Gamma\gg0.001M$, one should operate the muon collider
at its natural $R$ value of order $0.12\%$. The $\Delta\sigma_M$-induced
errors will be very tiny, both because $\Gamma/\sigma_M$ is very large
and because $\Delta\sigma_M/\sigma_M$ should be small for such $R$.
(b)
For a resonance with $\Gamma<0.001M$, one will typically wish to operate
at an $R$ that is significantly
smaller than that value which would yield equal statistical and systematic 
errors. In a typical case, this would mean a value of $R$ such
that $\Gamma/\sigma_M$ is larger than $5\div 10$.
Of course, if the resonance is extremely narrow,
it may happen that $\Gamma/\sigm$ is of order, or not much larger than, unity
even for $R=0.003\%$. In this case, it will normally
be essential to run with $R=0.003\%$ even though this
$R$ yields the smallest machine luminosity.  Larger values of $R$ lead
to a drastic decline in the signal to background ratio in a typical final
state that, in turn, leads to very poor statistical errors (given
the rather slow compensating increase with $R$ of the instantaneous luminosity).

For a narrow resonance
with $\Gamma\ll 0.001M$, the technique in which one sits on the resonance peak
and measures the cross section for two different values of
$\sigma_M$ ($\sigma_M^{\rm max}>\Gamma$ and $\sigma_M^{\rm min}<\Gamma$) 
is a strong competitor to the scan technique. $\Gamma/\sigma_M^{\rm central}$ 
(where $\sigma_M^{\rm central}=\sqrt{\sigma_M^{\rm max}\sigma_M^{\rm min}}$)
is determined by the ratio $r_c=\sigma_c(\sigma_M^{\rm
min})/\sigma_c(\sigma_M^{\rm max})$, where $\sigma_c$ is the
measured cross section.
For $\sigma_M^{\rm max}/\sigma_M^{\rm min}$ of order 5 to 20,
the statistical error in $\Gamma/\sigma_M^{\rm central}$ is smallest
if $\Gamma/\sigma_M^{\rm central}\sim 2\div 3$. The larger 
$\sigma_M^{\rm max}/\sigma_M^{\rm min}$, the smaller the 
statistical error for $\Gamma/\sigma_M^{\rm central}$. 
For a typical choice of $\sigma_M^{\rm max}/\sigma_M^{\rm min}=10$,
one finds a statistical error of $\Delta\Gamma/\Gamma\sim
1.8\Delta r_c/r_c$ for $\Gamma/\sigma_M^{\rm central}\sim 2\div 3$. 
This technique has the advantage that the           
$\Delta\sigma_M$-induced systematic error in $\Gamma$ is simply
given by $\Delta\Gamma/\Gamma=\Delta\sigma_M/\sigma_M$, while there
is no systematic error in the determination of any of the 
$B\equiv B_{\mupmum}B_F$
branching ratio products ($F$ = a particular final state).
Of course, if the resonance
is very narrow (\eg\ as narrow as a light SM Higgs boson
or a light pseudogoldstone boson), $\Gamma/\sigm^{\rm central}\sim 2\div 3$
will not be achievable. In this case, the best that one can do is
to employ $\sigma_M^{\rm min}$ ($\sigma_M^{\rm max}$)
as given by $R=0.003\%$ ($R=0.03\%$). The statistical error in $\Gamma$
for such a situation is typically still very good.

Let us now summarize how these results apply in the specific
cases we explored.

In the case of a three-point scan
of the SM Higgs boson, we have shown that in the region
$M_H\lsim 110\gev$ it is mandatory to have
$R\lsim 0.003\%$. In fact, even for
this very small $R$ value, $\Gamma_H/\sigma_M$ is still $\lsim 1$,
the very minimum needed for accurate measurements of resonance parameters.
For $\Gamma_H/\sigma_M\sim 1$ the fractional
systematic errors induced in $\Gamma_H$ 
from uncertainty in the beam energy spread are of order 
$3\Delta\sigma_M/\sigma_M$
for a $k=2$ scan. This should be compared
to the typically-expected statistical errors tabulated
in Table~\ref{fmcerrors}. For example, for $M_H=110\gev$
the statistical error in $\Delta\Gamma_H/\Gamma_H$ is $\sim 5\%$ 
for optimistic 4-year integrated luminosity of $L=4\fbi$ at $R=0.003\%$;
$\Delta\sigm/\sigm\lsim 0.01$ would be needed for the systematic
error to be smaller than the statistical error.
For the pessimistic 4-year integrated luminosity of $L=0.4\fbi$, the
statistical error would be much larger (\eg\ $\Delta\gamh/\gamh\sim 16\%$
at $\mh=110\gev$) and the $\Delta\sigma_M$-induced error would 
be much smaller than the statistical error if $\Delta\sigm/\sigm\lsim 0.01$.
Note, however, that increasing $R$ is not appropriate
as this would push one into the $\Gamma_H/\sigma_M<1$ region,
implying large statistical errors and still larger systematic
errors. 

For the $\sigma_M^{\rm max,min}$ on-peak ratio technique, one must
choose $\sigma_M^{\rm min}$ corresponding to $R=0.003\%$ 
($\Gamma_H/\sigma_M^{\rm min}\sim 1$). 
Results for statistical errors were presented in Table~\ref{fmcerrorsrc}.
As a point of comparison, for optimistic (pessimistic)
instantaneous luminosity and 4 years of operation, the net production rate
error after summing over important channels is
of order 0.8\% (2.7\%) for $M_H=110\gev$, and the
$\Delta\Gamma_H/\Gamma_H$ statistical error
is of order 8\% (25\%). Although the statistical $\Delta\sigma_c/\sigma_c$
fractional error is somewhat smaller for the ratio technique than
for the scan technique, the $\Delta\gamh/\gamh$ statistical error is larger.
However, the ratio technique might still be better if $\Delta\sigm/\sigm$
were as large as $5\%$, especially 
if the optimistic luminosity level is available.
This is because the systematic error in $\gamh$ is equal
to $\Delta\sigm/\sigm$ for the ratio technique as opposed to
$3\Delta\sigm/\sigm$ for the scan technique.
The $r_c$-ratio technique becomes increasingly superior as the assumed
luminosity increases. For $\mh\geq 130\gev$, the ratio technique gives
smaller statistical errors for both $\sigma_cB$ and $\gamh$
than does the scan technique (for which statistical
errors rapidly become very large). Indeed, the two procedures are nicely
complementary in that at least one of them will allow a measurement
of $\gamh$ with statistical accuracy below 6\% (20\%) for optimistic
(pessimistic) luminosity.

For the lightest PNGB ($P^0$) of an extended technicolor model, the
$\Delta\sigma_M$-induced errors for the three-point
scan method can be kept smaller than in the case
of the SM Higgs boson. This is because, for typical model
parameters, the $P^0$ has a width that is larger
than that of a SM Higgs boson; $\Gamma_{P^0}/\sigma_M>2$
is quite likely for $R\approx 0.003\%$. For example,
the parameter choices of \cite{mumu} give
$\Gamma_{P^0}\sim 5\mev$ vs. $\sigma_M\sim 2\mev$ at $M_{P^0}=100\gev$.
As we have seen, the resulting $\Gamma_{P^0}/\sigma_M\sim 2.5$ yields
$\Delta\sigma_M$-induced resonance parameter fractional errors of order
$\Delta\sigma_M/\sigma_M$. This can be compared to the statistical errors
given in Table~\ref{fmcerrorspgb}, computed assuming the pessimistic
4-year integrated luminosity of $L=0.4\fbi$.  
For example, at $\mpzero=110\gev$
the fractional statistical error for $\gampzero$ would be $\sim 0.04$,
which is much larger than the systematic error
if $\Delta\sigma_M/\sigma_M\sim 0.01$.
For $\mpzero <80\gev$, the statistical error in $\gampzero$ declines to the
$1\%\div 3\%$ level while the systematic error for
$\Delta\sigma_M/\sigma_M=0.01$ rises to about this same level.
As $\mpzero$ increases from $150\gev$ to $200\gev$, the statistical
error for $\gampzero$ rises from $\sim 5\%$ to $\sim 10\%$ while
the systematic error is below $1\%$ if $\Delta\sigma_M/\sigma_M=0.01$.
For optimistic integrated luminosity of $L=4\fbi$, the statistical
errors would be smaller than quoted above. For $\Delta\sigma_M/\sigma_M=0.01$,
the induced systematic errors would generally dominate for $\mpzero\leq
80\gev$. 

The $\pzero$ resonance is sufficiently narrow that one should also
consider using the $\sigma_M^{\rm max,min}$ on-peak ratio technique
to determine $\gampzero$.
For 4-year pessimistic luminosity operation we find the statistical errors
for $\gampzero$ given in Table~\ref{fmcerrorspgb}.
For the on-peak ratio technique, the systematic error in $\gampzero$
is equal to $\Delta\sigma_M/\sigma_M$ and for lower $\mpzero$ values
would only be smaller than the statistical $\gampzero$ error
if $\Delta\sigma_M/\sigma_M\lsim 0.01$.
For optimistic luminosity, the $1\%$ systematic error
induced in $\gampzero$ for $\Delta\sigma_M/\sigma_M=0.01$
would dominate over the statistical error for all but
$\mpzero\sim \mz$ and $\mpzero>150\gev$.
Most importantly, the statistical and systematic errors
of the ratio technique are at least as good as, and often
better than, obtained using the scan technique.
For $\mpzero\leq 110\gev$, the statistical $\gampzero$ errors
of the ratio technique are almost the same as obtained 
via the three-point scan (performed with $R=0.003\%$),
while the systemic $\gampzero$ errors from the ratio technique
are smaller ($\Delta\sigm/\sigm$ vs. $\sim 1.5\Delta\sigm/\sigm$).
For $\mpzero\geq 120\gev$, the $\Delta\sigm$-induced systematic
errors are comparable for the two techniques, but the statistical
errors for the ratio technique are substantially smaller than
for the three-point scan. Precision measurements
of the properties of a $\pzero$ resonance would, thus, always be best
performed using the ratio procedure.

In the case of the two resonances of the Degenerate BESS
model, we have determined (for several typical $R$ values) the region
of the model parameter space for which the
fractional errors in the resonance properties ($\Gamma$, $\ldots$)
induced by $\Delta\sigma_M$
are less than a given fixed value. Induced errors are
small for large resonance masses. But, for any given choice of $R$,
as the resonance mass is decreased, while maintaining model parameter
choices such that the model is still consistent with precision experimental
data, there comes a point at which, even for
$\Delta\sigma_M/\sigma_M\sim 0.01$, the induced error
becomes large. For allowed model parameter choices yielding
a resonance mass below this, the resonance's properties
cannot be measured accurately.  The lowest masses 
of the resonances that correspond to 
precision-data-allowed BESS model parameters,
and for which $\Delta\sigma_M$-induced
errors in the measured resonance properties are small, decrease rapidly
with decreasing $R$.  As a result, the ability to achieve $R<0.1\%$
at a muon collider would be crucial for exploring the low-resonance-mass
portions of Degenerate BESS parameter space not currently excluded
by precision data.

Finally, we have performed the analysis of two nearly degenerate
resonances, a situation encountered in a number of theoretical examples,
including the Degenerate BESS model and the minimal supersymmetric model.
We focused on the case in which the
total widths of the resonances are much smaller than the mass
splitting. We have shown that, in general, the
$\Delta\sigma_M$-induced fractional error in
the measured mass splitting, $M^s$, and in the leptonic
partial widths of the resonances
(or leptonic partial widths times final state branching fraction)
depend only on $M^s/\sigma_M$ and on the ratio of the two
partial widths, $a$. The main result is that the errors are
generally big for $M^s/\sigma_M$ less than a certain critical value
(typically in the $2\div 3$ range) that is a function of $a$. 
As $M^s/\sigma_M$ increases beyond
the critical value, the $\Delta\sigma_M$-induced fractional error in the
mass splitting approaches zero rapidly, whereas the fractional errors in the
partial widths approach $\Delta\sigma_M/\sigma_M$.  As a
concrete case, we have discussed the application to the
two spin-1 resonances of the Degenerate
BESS model for beam energy spread of the same order as the mass splitting
between them. We determined
the regions of the model parameter space in which $M^s$ could
be measured with $\Delta\sigma_M$-induced fractional error below a given
fixed value assuming a given value of $R$. The smaller the value
of $R$ that can be used while maintaining sufficient luminosity
for small statistical errors,
the larger the fraction of allowed parameter space for which $M^s$ can
be measured with small $\Delta\sigma_M$-induced error. In particular,
$R<0.1\%$ is required if we are to be able to 
separate the degenerate resonance peaks for model parameter
choices not already excluded by precision experimental data
in which the resonance masses are as low as $100\gev$.

\bigskip

\noindent{\bf Acknowledgements}
We wish to thank A. Blondel for the interesting discussions
during the CERN Workshop on muon colliders, where this paper was
conceived. JFG is supported in part by
the U.S. Department of Energy and by the Davis Institute for High
Energy Physics.

\clearpage

\clearpage

\clearpage

\clearpage

\clearpage

\clearpage

\clearpage

\clearpage

\clearpage

\clearpage

\clearpage

\clearpage

\clearpage

\clearpage

\clearpage

\clearpage

\end{document}